\documentclass[a4paper,usenatbib]{mn2e}
\usepackage{float}
\restylefloat{table}
\usepackage{times}
\usepackage{natbib}
\makeatletter
\renewcommand\hyper@natlinkbreak[2]{#1}
\makeatother
\usepackage{natbib,hyperref,ifthen} 
\usepackage{hyperref}
\usepackage{etoolbox}
\hypersetup{
 colorlinks=true,
 citecolor=blue}
\usepackage{caption}
\usepackage{subfigure}
 \usepackage{graphicx}
 \usepackage{graphics}
 \usepackage{amsmath}
 \usepackage{amssymb}		
  \usepackage{lipsum} 
 \usepackage{afterpage}
\usepackage{threeparttable}   
\usepackage{placeins}   

 \newcommand{\aap}{A\&A}

\newcommand{\apjs}{ApJS}
\newcommand{\apjl}{ApJL}

\newcommand{\hi}{\mbox{H{\sc i}}}

\newcommand{\kms}{km s$^{-1}$}
\newcommand{\mkms}{{\rm km\,s^{-1}}}

\newcommand{\Mo}{\rm M_{\odot}}

\newcommand{\dg}{^{\circ}}
\newcommand{\mjybeam}{\rm mJy\,beam^{-1}}

\newcommand{\acm}{\rm atoms\,cm^{-2}}

\newcommand{\marc}{mag\,arcsec$^{-2}$}

\newcommand{\e}[1]{\times 10^{#1}}
\newcommand{\HI}{\textsc{Hi}}
\newcommand{\nhi}{N_\HI}

\newcommand{\miriad}{\textsc{miriad}}
\newcommand{\casa}{\textsc{casa}}

 \renewcommand\appendix{\par
  \setcounter{section}{0}
  \setcounter{subsection}{0}
  \setcounter{figure}{0}
  \setcounter{table}{0}
  \renewcommand\thesection{ \Alph{section}}
  \renewcommand\thefigure{\Alph{section}\arabic{figure}}
  \renewcommand\thetable{\Alph{section}\arabic{table}}
}

\title[Observations of Virgo Southern Filament]{$\hi$ observations of galaxies in the southern filament of the Virgo \\
Cluster with the SKA Pathfinder KAT-7 and the WSRT}

\author[Sorgho et al.]{A. Sorgho$^{1}$\thanks{amid.ast@gmail.com}, K. Hess$^{1,2,3}$, C. Carignan$^{1,4}$, T. A. Oosterloo$^{2,3}$\\
\\
$^{1}$ Department of Astronomy, University of Cape Town, Private Bag X3, Rondebosch 7701, South Africa\\ 
$^{2}$ Netherlands Institute for Radio Astronomy (ASTRON), Postbus 2, 7990 AA Dwingeloo, The Netherlands\\
$^{3}$ Kapteyn Astronomical Institute, University of Groningen, PO Box 800, 9700 AV Groningen, The Netherlands\\
$^{4}$ Observatoire d'Astrophysique de l'Universit\'{e} de Ouagadougou (ODAUO), Ouagadougou, Burkina Faso}

\begin{document}

\date{}

\label{firstpage}
\maketitle

\begin{abstract}
We map the \hi\, distribution of galaxies in a $\sim 1.5^\circ \times 2.5^\circ$ region located at the virial radius south of the Virgo cluster using the KAT$-$7 and the WSRT interferometers. Because of the different beam sizes of the two telescopes, a similar column density sensitivity of $\nhi\sim 1\e{18}\,\acm$ was reached with the two observations over 16.5 \kms. We pioneer a new approach to combine the observations and take advantage of their sensitivity to both the large and small scale structures. Out to an unprecedented extent, we detect an \HI\, tail of $\sim 60$ kpc being stripped off NGC 4424, a peculiar spiral galaxy. The properties of the galaxy, together with the shape of the tail, suggest that NGC 4424 is a post-merger galaxy undergoing a ram pressure stripping as it falls towards the centre of the Virgo Cluster. We detect a total of 14 galaxies and 3 \hi\, clouds lacking optical counterparts. One of the clouds is a new detection with an \hi\, mass of $7\e{7}\, \Mo$ and a strong \hi\, profile with $W_{50} = 73$ \kms. We find that 10 out of the 14 galaxies present \hi\, deficiencies not higher than those of the cluster's late spirals, suggesting that the environmental effects are not more pronounced in the region than elsewhere in the cluster.
\end{abstract}



\section{Introduction}

When a galaxy resides in a dense region such as a cluster or a group, it undergoes different mechanisms that determine its morphology \citep[e.g.][]{Dressler1980} and gas properties \citep{Cayatte1990}. Galaxies evolving in dense environments tend to be \HI\, deficient with respect to their morphological counterparts evolving in less dense regions \citep{Haynes1984}.
More specifically in clusters of galaxies, the small scale structure interstellar medium (ISM) of spiral galaxies interacts with the large scale structure intracluster medium (ICM) through hydrodynamical processes that then drive the evolution of those galaxies in clusters. Perhaps the most important mechanism at work is the ram pressure (RP) stripping \citep[e.g.][]{Gunn1972,Vollmer2001}. It occurs when a galaxy moves through the ICM of a galaxy cluster and the external force felt by the ISM is greater than the gravitational restoring force from the galaxy disk.  It is known to be partly responsible for the \HI\, deficiency \citep{Bothun1982,Giovanelli1985,Vollmer2001} and star formation quenching \citep[e.g.][]{Balogh2000}. Typical examples of galaxies in the Virgo cluster undergoing RP stripping include NGC 4388 \citep{Oosterloo2005}, NGC 4522 \citep{Kenney2004a,Vollmer2004}, NGC 4402 \citep{Crowl2005} and NGC 4438 \citep{Chemin2005}.
Although this effect is strongest in the central regions of clusters, cosmological simulations reveal that it remains an important mechanism of gas stripping of galaxies out to the virial radius of clusters \citep{Tonnesen2007}.

While the optical characteristics of galaxies in dense environments seem well understood, we are still far from a complete picture of galaxy evolution.
Observations of the neutral atomic gas in dense environments are key to our understanding of the growth and depletion of gas disks, as well as star formation quenching in galaxies.  Further, they provide tests for numerical simulations and constraints for modelling galaxy evolution \citep[e.g.][]{Dave2013}. In particular the neutral gas provides an opportunity to map the outer regions of galaxies and their surrounding environment, allowing us to directly probe their interactions with the IGM and with their neighbours.


To date, about 1300 galaxies have been identified, based on their morphology and radial velocities, as true members of the Virgo Cluster \citep[e.g.][]{Sandage1985,Binggeli1987}.  
Located at 16.8 Mpc \citep[distance of M87,][]{Neill2014}, the recession velocity of the cluster is $\sim1050$ \kms, and its velocity dispersion is $\sim700$ \kms\, \citep{Binggeli1993}. Detailed X-ray observations of the hot cluster gas \citep{Bohringer1994} revealed several substructures, suggesting that the cluster is not virialised, as seen in other clusters \citep[see e.g. for the Coma \& Antlia clusters,][]{Briel1992, Hess2015a}. The Virgo Cluster, instead, is still growing. Subclusters around the ellipticals M86 and M49 \citep[whose systemic velocities are respectively -181 and 950 \kms,][]{Kim2014} are merging with the main cluster around the central elliptical M87. Several \hi\, imaging surveys of the cluster revealed that the \hi\, disks of its central late-type galaxies are truncated to well within the optical disks, suggesting that ICM - ISM interactions play an important role in driving the evolution of galaxies in the inner region of the cluster \citep{Warmels1988a, Cayatte1990, Chung2009, Boselli2014}. Wide-area \hi\, surveys like HIPASS \citep[\hi\, Parkes All-Sky Survey,][]{Barnes2001}, ALFALFA \citep[Arecibo Legacy Fast ALFA,][]{Giovanelli2005} and AGES \citep[Arecibo Galaxy Environment Survey,][]{Auld2006} done with single-dish telescopes have helped detect many low \hi\, mass galaxies such as early-type and late-type dwarf galaxies, but also starless ``dark" galaxies and several \hi\, clouds \citep[e.g.,][]{Alighieri2007, Kent2007}. In particular, ALFALFA observations allowed the detection of complex structures associated with Virgo galaxies \citep[e.g.][]{Koopmann2008}, showing probable signs of interactions in the cluster.

\begin{figure}
	\hspace{-50pt}
	\includegraphics[width=1.5\columnwidth]{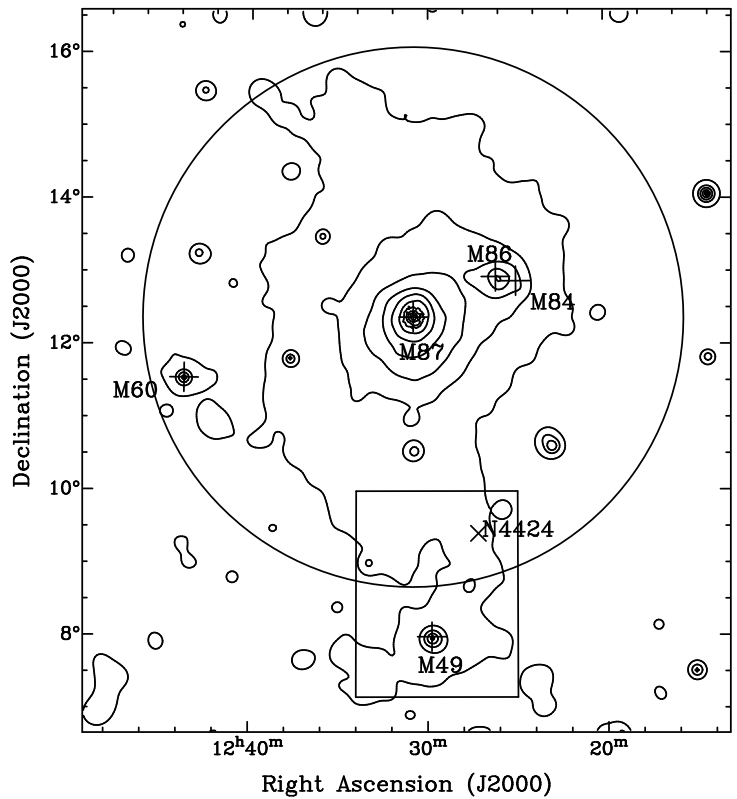}
	\caption{X-ray map of the Virgo Cluster \citep[$0.5-2.0$ keV, {\it ROSAT};][]{Bohringer1994}. The circle shows the virial radius of the cluster \citep[$\sim3.7\dg$ or 1.08 Mpc,][]{Arnaud2005, Urban2011a} around M87, and the box is the region observed in the present work. The labelled plus signs show the major ellipticals of the cluster, and the cross indicates the position of the \HI-tailed galaxy NGC 4424.}\label{fig:virgo-rosat}
\end{figure}

We use the KAT-7 and the WSRT to study galaxies located in an X-ray filament to the South-West of the cluster connecting the main cluster with the substructure associated with M49. We observe a $\sim 1.5^\circ \times 2.5^\circ$ region in the M49 subcluster. It is located south of the Virgo Cluster, $\sim3\dg$ away from the elliptical M87, and contains the elliptical M49 (see Fig. \ref{fig:virgo-rosat}). The field also contains the one-sided \hi\, tail galaxy, NGC 4424, previously observed using the VLA \citep{Chung2007,Chung2009}. Taking advantage of the short baselines of the KAT-7 telescope sensitive to extended structures and the high spatial resolution of the WSRT telescope, we map late-type galaxies in the region down to low column densities. Given the large difference in the resolutions of the two arrays, we develop a new technique to combine the data from the two telescopes.

We begin with a description of the observations with the two telescopes, as well as the technique adopted for the data combination in section \ref{sec:ODR}. The results are presented in section \ref{sec:res} where the \hi\, map of the region as well as the \hi\, parameters of the detections are given. In section \ref{sec:dis} we discuss the results in light of the cluster environment effects, and give a summary in section \ref{sec:summary}. The contours and \hi\, profiles of all the detections are presented in the Appendix.

\section{Observations and Data Reduction}\label{sec:ODR}


\subsection{KAT-7 observations and reduction}
The Karoo Array Telescope (KAT-7) observations were conducted between July 2013 and May 2014 during a total of 14 sessions. All observations were centred at a frequency of 1418 MHz and the correlator mode provided a total bandwidth of 25 MHz, covering the velocity range $-2115$ to 3179 \kms\, divided in 4096 channels being 6.1 kHz (1.28 \kms) wide.

The observation technique used with KAT-7 consists in alternating between the target and a phase calibrator (1254$+$116). The flux calibrator (either PKS 1934-638 or 3C 286) was observed two or three times per session and the phase calibrator every 20 min to calibrate the phase stability. The choice of these calibrators lies on their close angular distance from the target field and their good quality status as given by the ATCA Calibrator Database\footnote{http://www.narrabri.atnf.csiro.au/calibrators/} and the VLA Calibrator Manual\footnote{https://science.nrao.edu/facilities/vla/docs/manuals/cal}. The calibrator must be as close as possible to the target, and have a good quality rating in the 21cm L-band.

\begin{figure}
		\centering
		\includegraphics[width=\columnwidth]{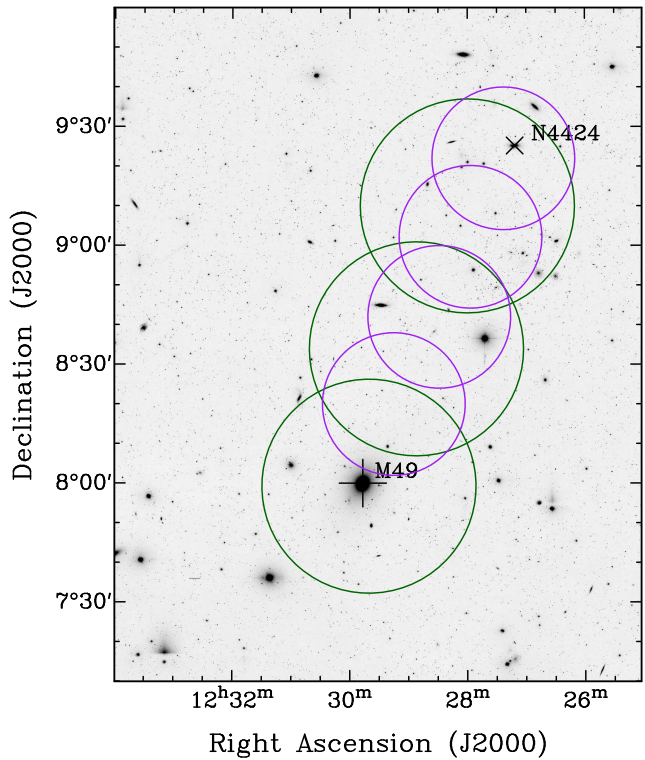}
\caption{The KAT-7 ({\it green}) and WSRT ({\it purple}) primary beams on an optical SDSS $r$-band grayscale, showing the pointing positions of the observations. The field is the same as the box in Figure \ref{fig:virgo-rosat}.}\label{fig:kat_wsrt-beams}

\end{figure}

We used three pointings to cover the desired field, uniformly separated (by $\sim72\%$ of the primary beam) and following the line connecting the objects NGC4424 and M49 (see Fig. \ref{fig:kat_wsrt-beams}). The upper pointing was observed for $\sim30$ hours split in 6 sessions, while the two others were given 24 hours each during a total of 8 sessions. Initially, the observations were focused on NGC 4424 and its \hi\, tail; however, to have a complete view of the tail and to probe the whole region between the galaxy and the elliptical M49, the upper pointing was expanded to a mosaic.

During these observations, KAT-7 was still in the commissioning phase and the RFI monitoring was still ongoing, therefore special care needed to be given to the inspection of the data beforehand. Manual flagging was performed to remove RFI, and special care was taken to suppress the effects of Galactic \hi\, on sources of interest. The standard \casa\, package \citep[version 4.1.0,][]{McMullin2007} was used to image the data. The observing sessions were individually reduced, and continuum subtraction was performed by using a first order fit over line free channels. Because KAT-7 does not use Doppler tracking and \casa\, 4.1.0 does not fully recognise frequency keywords, special care was taken to produce image cubes with the proper velocity coordinates \citep[see][]{Carignan2013a}. The parameters of the resulting primary beam-corrected data cube are summarised in Table \ref{tb:par-cubes}.

\begin{table}
	\begin{center}
  \caption{Summary of the KAT-7 and WSRT image cubes.\label{tb:par-cubes}}
  \begin{tabular}{ l | c | c }
      \hline\hline
      Parameter & KAT-7 & WSRT\\
      \hline
			Weighting function & Robust = 0 & Robust = 0.5 \\
			Channel width & 15.2 \kms & 16.5 \kms \\
			FWHM of beam & $4.4'\times3.1'$ & $2.9'\times0.6'$ \\
			PA of beam   & $\sim-8\dg$ & $\sim1\dg$ \\
			Map gridding & $55''\times55''$ & $5''\times5''$ \\
			Map size & $3.9\dg\times3.9\dg$ & $1.7\dg\times2.3\dg$ \\
			RMS noise & $2.5\,\mjybeam$ & $0.35\,\mjybeam$ \\
			$\nhi$ ($1\sigma$ \& 16.5 \kms) & 	$0.9\e{18}\,\rm cm^{-2}$	& $1.0\e{18}\,\rm cm^{-2}$	\\
			\hline 
  \end{tabular}
  \end{center}
\end{table}


\subsection{WSRT observations and reduction}\label{sec:wsrt-obs}
A total of 4 sessions of 12h observation each were conducted in April 2006 with the Westerbork Synthesis Radio Telescope (WSRT) single-feed system, prior to the installation of APERTIF (Aperture Tile In Focus). An individual pointing was observed during each session; two calibrators were also observed, 3C 147 and CTD 93. Unlike the KAT-7 observations, WSRT observations do not require gain calibrators at regular time intervals. This is because the phase of the WSRT system is quite stable, and the calibrators are observed only at the beginning and at the end of the individual 12h observing runs to fix the flux scale. To be more sensitive to extended features, the {\it Maxi-Short} configuration was adopted for the observations, and the centres of the individual pointings ($\sim22'$ apart) were chosen, like for the KAT-7 observations, such that the sensitivity is uniform across the field (see Fig. \ref{fig:kat_wsrt-beams}). The L-band correlator used provides a 20 MHz band split over 1024 channels, i.e, 4.12 \kms\, per channel.

We reduced the WSRT data using \miriad\, \citep{Sault1995}, and following the standard procedure for WSRT data reduction. Gain calibration was achieved by self-calibrating on background continuum sources throughout the observed field. Continuum subtraction was performed with the {\tt uvlin} task in the {\it u,v} plane and the solutions of the self-calibrated continuum data were subsequently applied to the calibrated line data. Next, the spectral regions of interest were selected and every 4 channels were averaged, changing the spectral resolution from 4.12 \kms\, to 16.49 \kms. The individual observations were then co-added into a single {\it u,v} dataset.

A mosaic datacube was produced using the {\it Robust} weighting with a robustness $R=0.5$, and a sampling interval (pixel size) of $5''$. A Gaussian taper of $30''$ was also applied. Because of the East-West layout of the WSRT array and the low declination of the field, the restoring beam is very elongated: $2.9'\times0.6'$. The datacube was then cleaned in \miriad\, using {\tt MOSSDI}, the cleaning task for mosaics. The cleaning was performed only in channels containing emission, and in regions around emission features. A primary beam correction was finally applied to the cube. The third column of Table \ref{tb:par-cubes} summarises the parameters of the WSRT image cube.


\subsection{Combination of the two datasets}
Data combination is widely used in radio astronomy, to get the most out of observations. Several methods have been developed, performing the combination in either the image domain or in the {\it u,v} plane \citep[see e.g.,][]{Stanimirovic2002}. Techniques of combining data from different telescopes are generally used for short spacing correction, i.e, used to combine single-dish data to interferometer data \citep[e.g.,][]{Vogel1984, Schwarz1991, Stanimirovic1999a, Kurono2009, Faridani2014}. As for the combination of two interferometer data from different arrays, the procedure is quite different \citep[see e.g.][]{Carignan1998}. The straightforward way of combining interferometer data from different arrays is adding them in the {\it u,v} plane. This works best when the individual data have the same sensitivity, which is not always the case.

While the average rms noise in the KAT-7 data is $\sim2.5\,\mjybeam$, that in the WSRT is much smaller, $\sim0.35\,\mjybeam$. Since the weight assigned to the datasets during combination goes as $1/\sigma_{\rm rms}^2$, the WSRT data end up having a much larger weight relative to the KAT-7 data. If $I_K,\,w_K$ are respectively the intensity and weight of a given pixel in the KAT-7 data and $I_W,\,w_W$ those of the corresponding pixel in the WSRT data, the intensity of the corresponding pixel in the resulting combined data is
\begin{equation}
I_c = {{I_K\,w_K + I_W\,w_W}\over{w_K + w_W}}.
\end{equation}
So, for $w \propto 1/\sigma_{\rm rms}^2$, $w_K \ll w_W$ and $I_c \sim I_W$.

Therefore, the traditional combination in the $u,v$ plane using the dedicated \miriad\, tasks \citep[see e.g][]{Hess2009} results in a largely WSRT feature-dominated data, with the KAT-7 features being essentially nonexistent. To avoid this, we performed the combination in the image plane based on column density sensitivity. This procedure provides a complex final beam that has a double gaussian shape. The ratio between the rms noises is more or less equal to the ratio between the respective synthesised beam areas. The column density of a source is given by
\begin{equation}\label{eq:coldens}
\nhi/\Delta v = 1.1\e{21}\,\left({I_\HI^{i,j} \over b_{\rm min}\,b_{\rm maj}}\right)
\end{equation}
or, in terms of brightness temperature,
\begin{equation}\label{eq:tb}
T_B = 611\,\left({I_\HI^{i,j} \over b_{\rm min}\,b_{\rm maj}}\right),
\end{equation}
where $b_{\rm min}$ and $b_{\rm maj}$ are respectively the minor and major FWHM of the synthesised beam in arcsec, and $I_\HI^{i,j}$ the \hi\, intensity of a pixel of coordinates ({\it i,j}) in $\mjybeam$. The brightness temperature $T_B$ is expressed in $K$. The column density is proportional to the flux and inversely proportional to the beam area. Therefore, the two datasets end up having the same order of column density sensitivity. The most appropriate way of combining them is by converting the cubes into units of column density per unit velocity in the image plane.

The continuum-subtracted KAT-7 {\it u,v} data was first exported from \casa\, to \miriad\, via UVFITS format. Once imported into \miriad, the velocity axis was resampled to match that of the WSRT data in velocity range and width. This was achieved with the {\tt UVAVER} task. The spatial centre of the mosaic was set to be that of the WSRT cube, and a gaussian taper of $30''$ was also applied. 
The resulting cube was cleaned and primary beam corrected using the {\tt MOSSDI} and {\tt LINMOS} tasks respectively. However, since the KAT-7 field of view is larger than that of the WSRT, we used {\tt REGRID} to regrid the spatial axes of the WSRT cube to the KAT-7 new cube.

At this stage, both image cubes have the same dimensions, coordinate grids, reference pixel and pixel size. This way, any pixel in one cube has the same world coordinates as its corresponding pixel in the other cube. Equation \ref{eq:coldens} was then used, with the {\tt MATHS} task, to convert them into column density cubes. Since the cubes initially had units of flux and Equation \ref{eq:coldens} uses the line flux, the output cubes have units of column density per unit velocity $\rm{cm^{-2}}\,(\mkms)^{-1}$ or, using Equation \ref{eq:tb}, units of brightness temperature $K$. The average rms noise levels (measured in the central regions of the cubes and over line-free channels) are 31 mK and 5 mK respectively in the output KAT-7 and WSRT data cubes: a weight ratio of 1.26.

\begin{figure}
		\hspace{-20pt}
		\includegraphics[width=1.25\columnwidth]{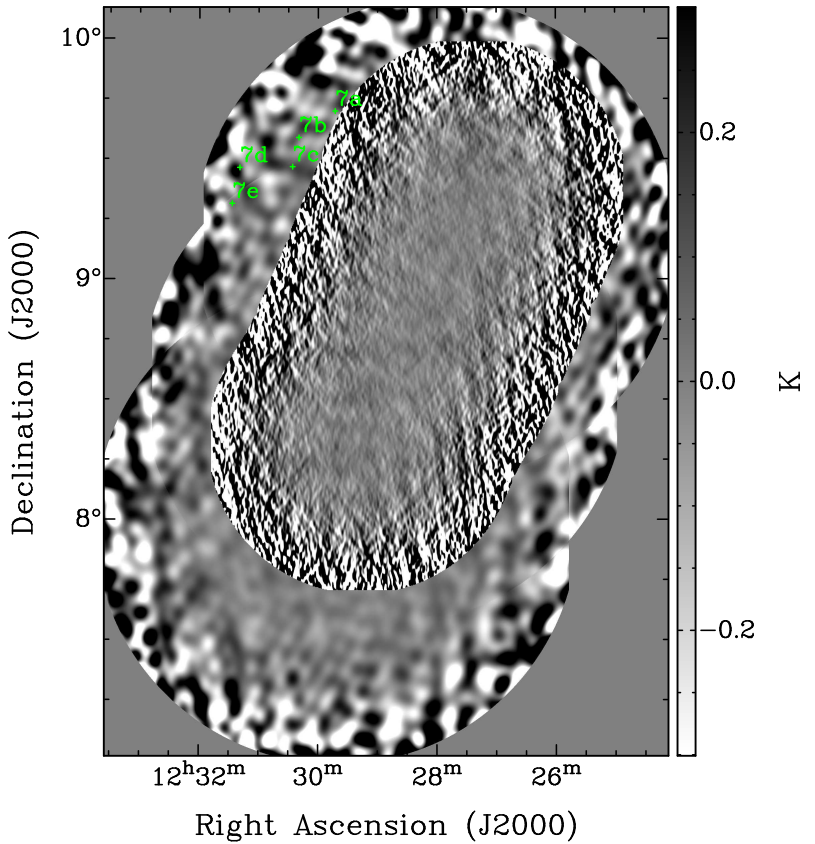}
		\caption{Map showing the variation of the noise in the combined cube. The five {\it green dots} show the positions of the \hi\, clouds detections reported in \citet{Kent2007} and discussed in section \ref{sec:gascloud}.}\label{fig:noise-var}
\end{figure}

Finally, the {\tt MATHS} task was used to linearly combine the two data cubes, with the expression
\begin{equation}\label{eq:data-comb}
I_c = {{1.26\,I_K + I_W}\over 2.26}.
\end{equation}
The rms noise in this final cube is 27 mK (or $7.9\e{17}\,\rm{cm^{-2}}$) at a velocity resolution of 16.5 \kms\, and allowed the detection of 12 objects at the $3\sigma$ level. In Figure \ref{fig:noise-var} we present a map of the noise in the combined data, showing the variation of the noise across the field.
To be sensitive to low surface brightness objects, the data cube was furthermore Hanning-smoothed to 33 \kms\, with an rms noise of 23 mK ($7.0\e{17}\,\rm{cm^{-2}}$), and 5 more objects were detected in this new cube.

To obtain the approximate size of the final beam in the combined data, we consider two individual two-dimensional gaussian functions of standard deviation $\sigma_\alpha$ and $\sigma_\delta$ each and along each axis, representing the beams of the initial datasets and whose FWHMs can each be written $\Delta\alpha = 2\sqrt{2\ln{2}}\,\sigma_\alpha$ and $\Delta\delta = 2\sqrt{2\ln{2}}\,\sigma_\delta$, respectively along the major and minor axes. Assuming that position angle of the two beams are equal (Table \ref{tb:par-cubes} shows that they are similar in the present case), we can sum the two gaussian functions and the major and minor axes of the resulting function can respectively be written $\Delta\alpha_c = [(0.56\,\Delta\alpha_K)^2 + (0.44\,\Delta\alpha_W)^2]^{0.5}$ and $\Delta\delta_c = [(0.56\,\Delta\delta_K)^2 + (0.44\,\Delta\delta_W)^2]^{0.5}$. Using the parameters of the individual beams in Table \ref{tb:par-cubes}, we obtain the approximate parameters of the combined beam to be $2.8'\times1.7'$.

\section{Results}\label{sec:res}


\subsection{\hi\, maps and \hi\, properties}

\begin{figure*}
\hspace{-100pt}
\includegraphics[width=1.2\textwidth]{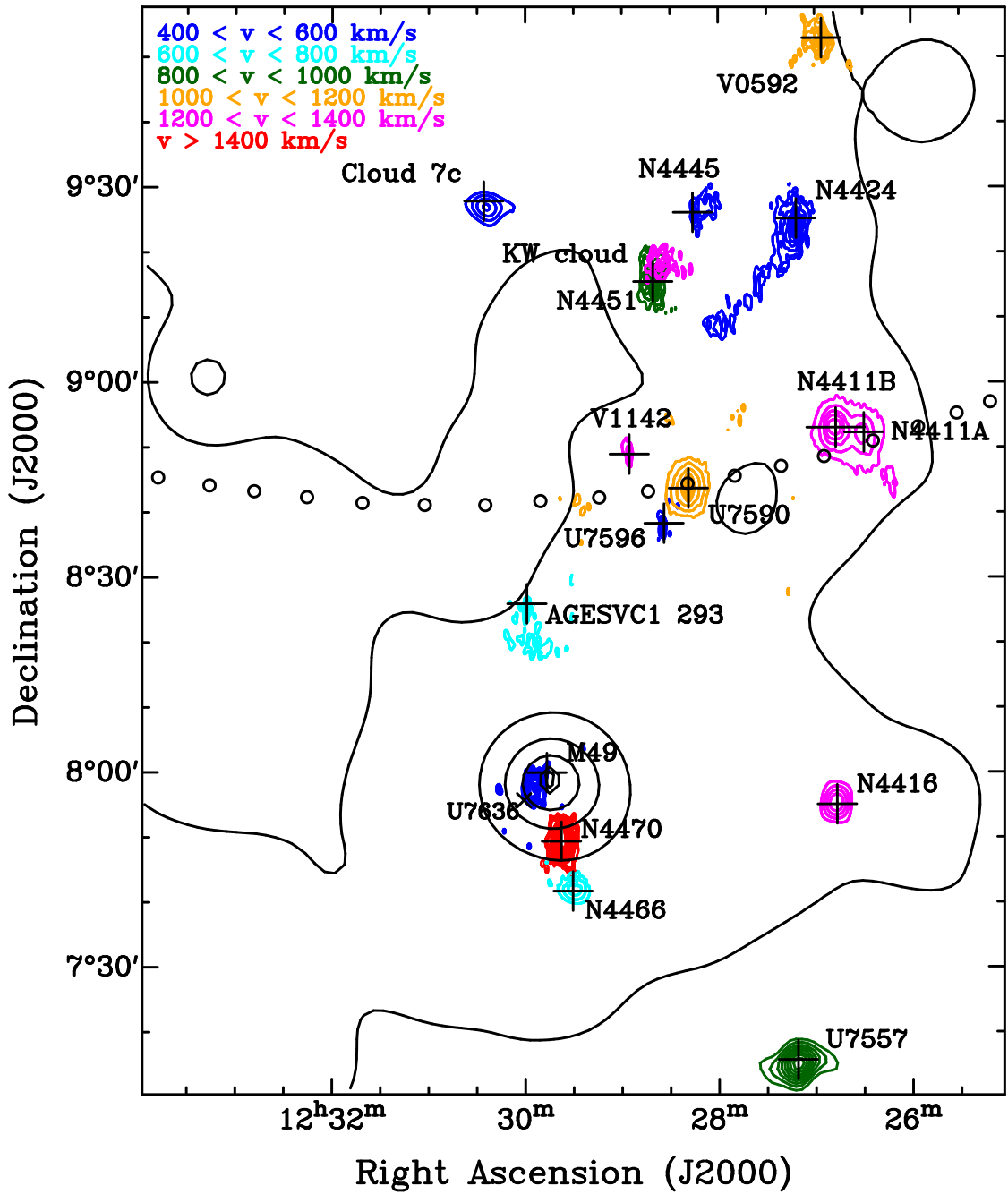}
\caption{The zoomed-in observed region (box of Fig. \ref{fig:virgo-rosat}). The black contours are the X-ray levels (4.2, 4.4, 4.7, 5.0, 5.2, 5.5 \& $5.8\e{-11}\,\rm ergs\,cm^{-2}\,s^{-1}$). The coloured contours are \HI\, column densities (as obtained from the combined map), ranging from $5\e{18}$ to $8\e{20}\,\rm cm^{-2}$ (see Figures \ref{app:n4424}-\ref{app:ages293} for individual contours levels) . For reference, the giant elliptical M49 is at $v_{\rm sys} = 950$ \kms. The dotted line shows the virial radius of the cluster.}\label{fig:rosat-hi}
\end{figure*}

Figure \ref{fig:rosat-hi} shows the \hi\, total intensity maps of all detections made in the field, overlaid on an X-ray map \citep[$0.5-2.0$ keV, {\it ROSAT};][]{Bohringer1994} of the cluster. A total of 17 objects were detected, including 14 galaxies with confirmed optical counterparts and 3 \hi\, clouds. The \hi\, detection of 2 of the 3 \hi\, clouds was previously reported, but they have not been optically detected as they have no counterparts in the SDSS. As for the third cloud, no \hi\, nor optical detection has been reported in the literature. We refer to this as the KW (KAT-7+WSRT) cloud. We discuss these individual clouds in section \ref{sec:gascloud}. In Table \ref{tb:maintable} we present the \hi\, properties of these different detections as determined from the combined image cube. The columns (1), (2) \& (3) contain respectively, the common names of their optical counterparts when available, their optical J2000 coordinates and morphological types when available. In columns (4) and (5) we list respectively the optical diameters at a surface brightness of 25 mag $\rm arcsec^{-2}$ in the B band, and inclinations as obtained from the RC3 catalogue \citep{Corwin1994}. In column (6) we list the Tully-Fisher distances used to estimate the \hi\, masses, mostly from \citet{Solanes2002}. In columns (7) and (8) we list, respectively, the systemic velocities and velocity widths obtained from the combined image cube. In columns (9) and (10) we give the measured \hi\, masses and deficiency parameters of the detections along with error estimates, and finally, in column (11) their projected angular distances from M87.

\subsection{NGC 4424 \hi\, tail}

\begin{figure}
\begin{minipage}[]{\linewidth}
	\centering
	\includegraphics[width=0.85\columnwidth]{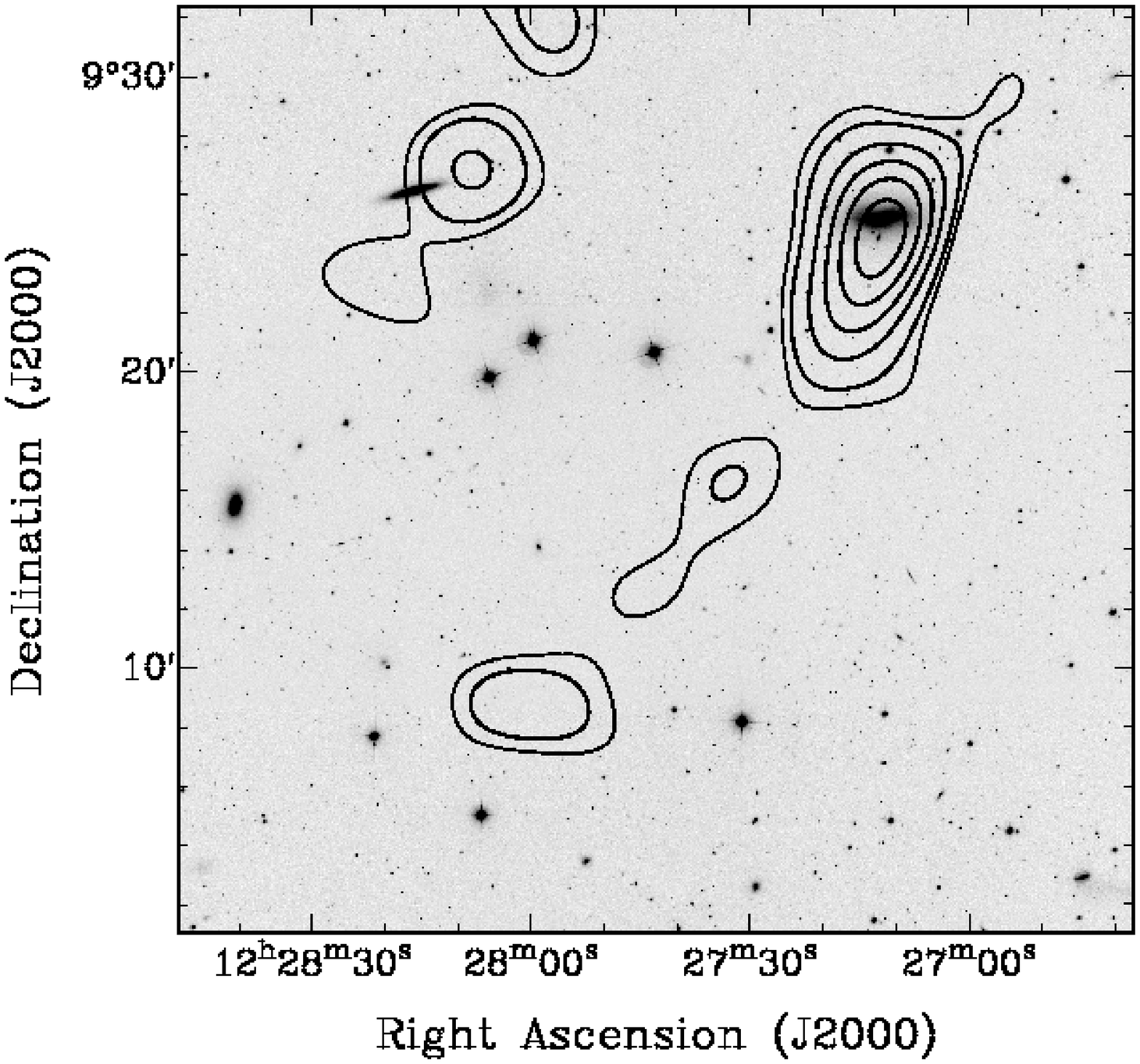}
\end{minipage}
\vspace{10pt}

\begin{minipage}[]{\linewidth}
	\centering
	\includegraphics[width=0.85\columnwidth]{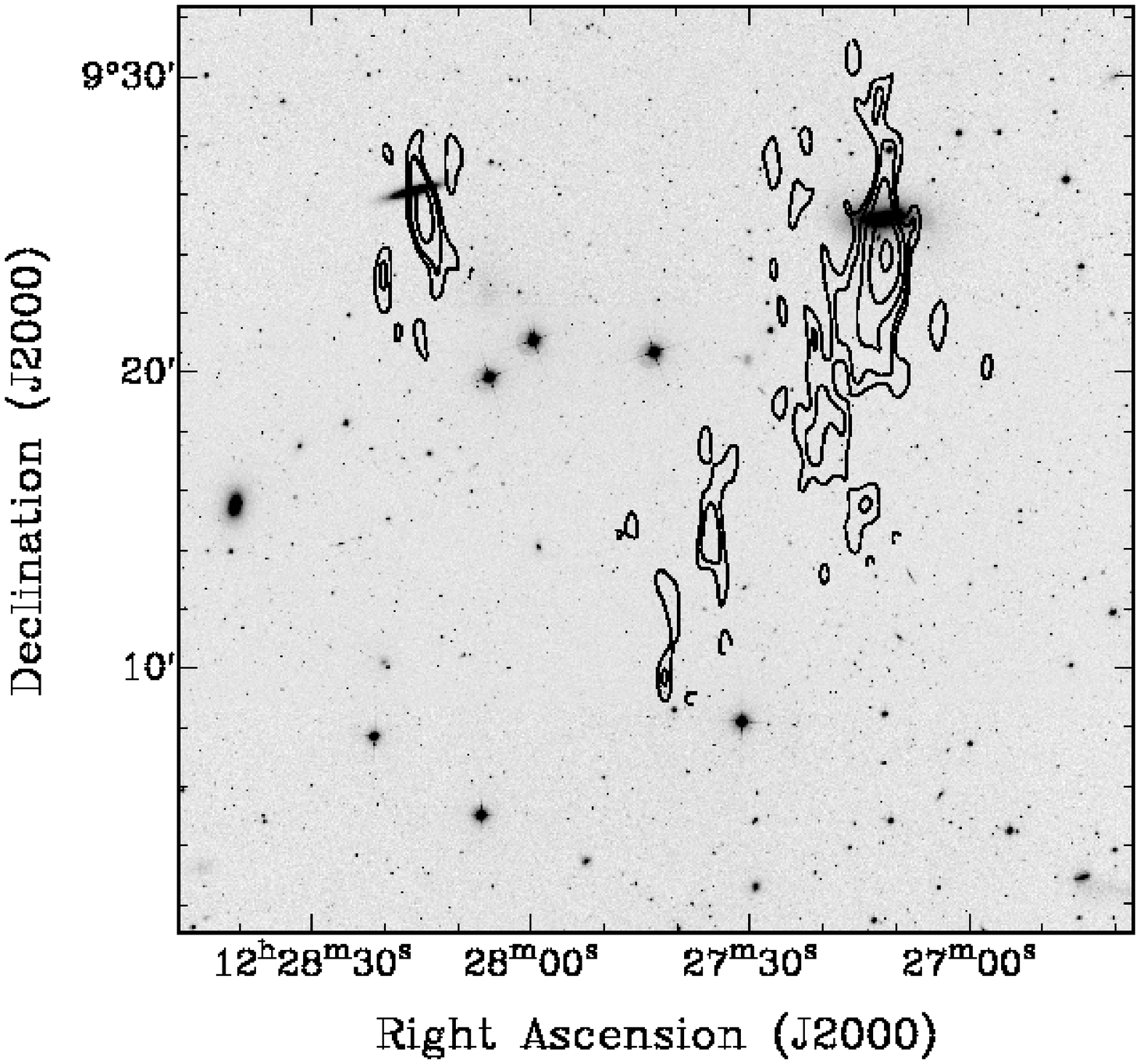}
\end{minipage}
\vspace{10pt}

\begin{minipage}[]{\linewidth}
	\centering
	\includegraphics[width=0.85\columnwidth]{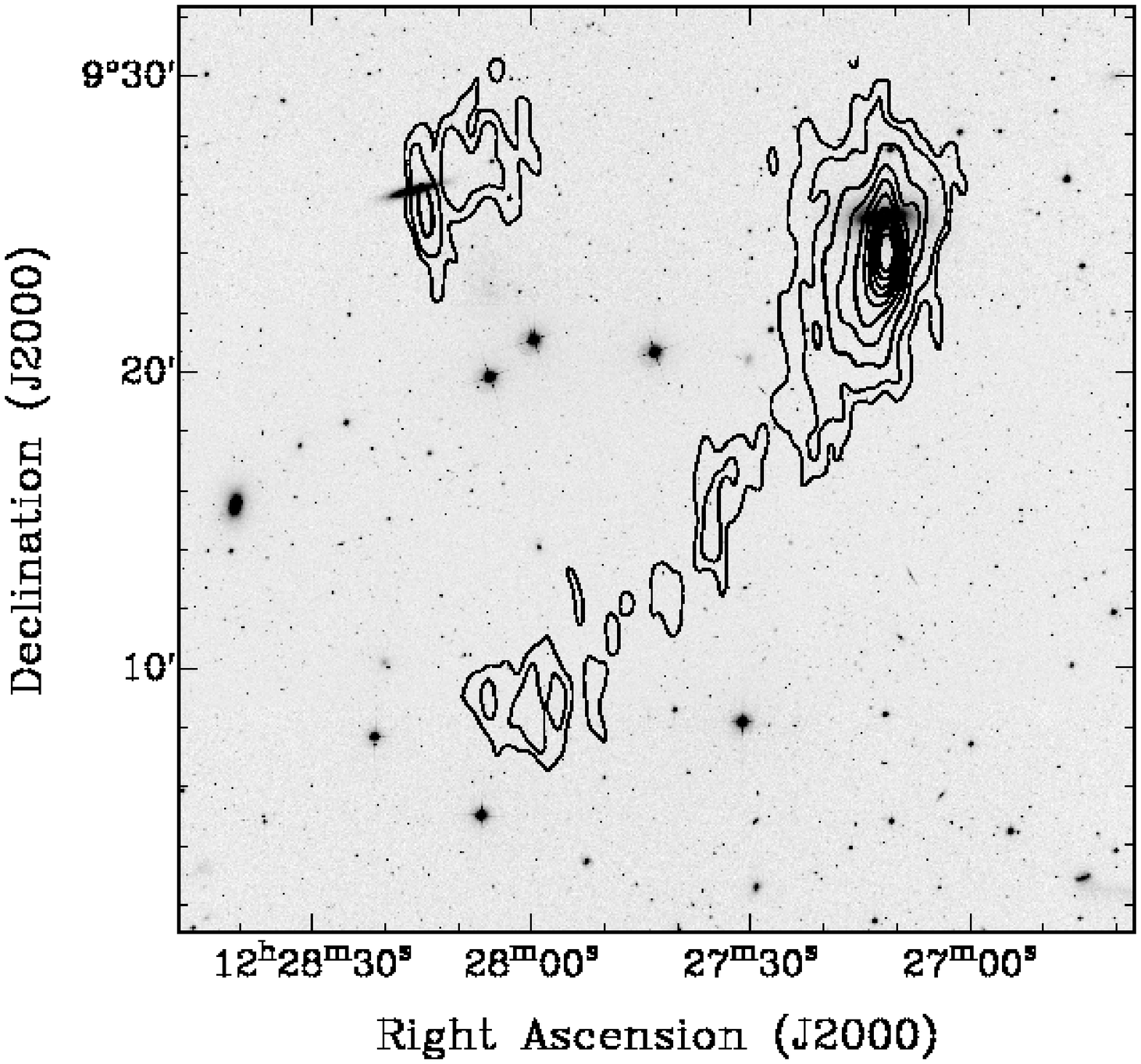}
\end{minipage}
\caption{NGC 4424 as seen in the KAT-7 ({\it top}), WSRT ({\it middle}) and combined ({\it bottom}) data, overlaid on SDSS r-band grayscale. The contour levels of the KAT-7 and WSRT maps are respectively $0.7\,(3\sigma)$, 1.0, 2.0, 3.0, 4.0, $5.0\e{19}\,\acm$ and 0.75, 1.5, 3.0, 6.0, 12.0, $15.0\e{19}\,\acm$, and those of the combined map are 0.5, 1, 2, 3 ... 10 $\times10^{19}\,\acm$.}\label{fig:n4424-comparison}
\end{figure}

The main goal of combining the two data sets was to achieve better \hi\, column density sensitivity and detect fainter structures. In the case of the galaxy NGC 4424, we seek to map the full extent of the \HI\, tail. In Fig. \ref{fig:n4424-comparison} we present a comparison of the \HI\, map of the galaxy as obtained in the three different datasets (KAT-7, WSRT \& combined). While the tail is seen in the KAT-7 data, its southern end is not detected in the WSRT data. When combined, the data reveal the structure of the tail with a better sensitivity of $\nhi \sim 8\e{17}\,\rm cm^{-2}$ over 16.5 \kms. Assuming a distance of 15.2 Mpc \citep{Cortes2008}, the tail extends out to $\sim60$ kpc from the main envelope, with an \hi\, mass of $(4.3\pm0.9)\e{7}\,\Mo$. This represents $\sim20\%$ of the total \hi\, mass of the galaxy. In previous VLA observations of the galaxy by \citet{Chung2007} down to $\sim 2\e{19}\,\rm cm^{-2}$, only $\sim20$ kpc of tail was seen. The present observations therefore reveal an \HI\, tail three times longer than previously observed.

\subsection{Gas clouds}\label{sec:gascloud}
Three \hi\, clouds were detected in the combined image cube: the M49 cloud spatially located near M49 and at a systemic velocity of 476 \kms, the KW cloud located at $v_{\rm sys} = 1270$ \kms\, and whose spatial coordinates coincide with NGC 4451 but with an offset of 400 \kms\, in velocity (see Figure \ref{fig:rosat-hi}), and Cloud 7c ({\it aka} AGC 226161), previously observed by \citet{Kent2007}.

The M49 cloud lies, both in velocity and space, between the late-type dwarf galaxy UGC 7636 (at 276 \kms) and the giant elliptical M49 (at 950 \kms). Previous \hi\, and optical observations of these galaxies revealed that the cloud is a result of the combined effect of gas stripping due to RP and tidal interaction between the dwarf galaxy and the elliptical \citep{Henning1993, McNamara1994}. The gas cloud would have been stripped from the ISM of the dwarf galaxy due to M49. Later multi-wavelength observations by \citet{ArrigoniBattaia2012} confirmed these results. It was also detected in the AGES Survey \citep{Taylor2012}.

As for the KW cloud, no previous detection was reported in the literature, and follow-up observations are needed to confirm the lack of a stellar counterpart. The cloud has an \hi\, mass of $(7\pm1)\e{7}\,\Mo$ at the distance of M87, and presents a Gaussian profile of width $W_{50} = 73.2$ \kms.

The cloud 7c, detected with $M_\HI=(6\pm1)\e{7}\,\Mo$ (assuming a distance of M87) and a velocity width of 74 \kms, is one of the 8 objects that \citet{Kent2007} reported to not have stellar counterparts, and qualified as ``optically unseen detections''. The authors, who reported an \hi\, mass of the cloud almost 3 times higher than what is measured here, argue that 7c is part of a complex of 5 \hi\, clouds -- {\it cloud 7} -- among which only 2 (7c and 7d) are confirmed by follow-up VLA observations \citep{Kent2010}. The authors, however, explain that the remaining 3 objects were not expected to be confirmed by the follow-up observations because either distant from the pointing centre and therefore below the $3\sigma$ detection limit (7b \& 7e) or outside the field of view (7a). At the positions of the clouds 7b and 7c, the noise level in the present observations is about 89 mK (or $\rm 1.6\e{17}\,cm^{-2}(\mkms)^{-1}$), which gives a $3\sigma$ detection limit of $1.25\e{6}\, \Mo/\mkms$. The clouds 7d and 7e are located at the edge of the map, and the detection limit at their position increases to $3.2\e{6}\, \Mo/\mkms$. At the position of 7a which coincides with the edge of the WSRT beam (see Figure \ref{fig:noise-var}), the detection limit goes up to $1.1\e{7}\, \Mo/\mkms$. The ALFALFA \hi\, masses reported in \citet{Kent2007} for the five clouds put 7a, 7d and 7e below the detection limits reported above, even if one adopts a velocity width as low as 50 \kms. However, the 7b cloud is well above the detection limit and it comes as a surprise that the cloud was not detected. This fact, along with the low detected mass of 7c, suggests that some low column density gas is being missed in the region.

Although not contained in any major catalogue, a matching optical redshift for AGESVC1 293 was previously associated with an optical counterpart which has an optical redshift in the SDSS database \citep{Taylor2012}. We therefore do not classify the object as an \hi\, cloud, but do not include it in the analysis in section \ref{sec:gas-content} since its optical properties are not clearly determined.


\section{Discussion}\label{sec:dis}


\subsection{\hi\, morphologies}\label{sec:hi-morph}
The Virgo Cluster is not dynamically relaxed \citep{Binggeli1987, Binggeli1993}, but contains different substructures with the largest being the subclusters A (centred on M87) and B (centred on M49). The ensemble of late type galaxies in the cluster presents a larger velocity dispersion than early types, suggesting that many of the late types are in the stage of infall \citep{Tully1984, Binggeli1987}. 

It is well known that the infall of the galaxies in a cluster is likely to happen along filaments \citep[e.g.,][]{Porter2007, Mahajan2012}. Furthermore, the distribution of the hot X-ray in the Virgo cluster (see Fig. \ref{fig:virgo-rosat}) suggests a filamentary structure connecting the subclusters A and B, and the distribution of the detected galaxies in the region seems to follow this filament. 
In Fig. \ref{fig:rosat-hi} we present the observed region of Fig. \ref{fig:virgo-rosat}, showing an overlay of the \hi\, maps of detections on the hot X-ray gas in the region. Most of the detections belonging (spatially) to the filament present an asymmetry to the south in their \hi\, morphologies; these are NGC 4424, NGC 4451, NGC 4445, UGC 7596 and AGESVC1 293. All these galaxies have systemic velocities comprised between 400 and 900 \kms, and are blueshifted with respect to the systemic velocity of the cluster ($\sim1050$ \kms). They seem to be falling inside the cluster from the back, along the filament.

Previous works showed that RP causes cluster galaxies to begin forming \hi\, tails at about the virial radius \citep[e.g.,][]{Chung2007}. Fig. \ref{fig:rosat-hi} shows that most of the detections that are projected in the filament are located about a virial radius from M87, and might well be subject to RP stripping. However, except for NGC 4424, none of them exhibit obvious \hi\, tails.

If the asymmetry observed in the galaxies is a result of RP stripping due to the relative movement of the galaxies through the hot halo of the main cluster of Virgo, one should expect a correlation between the asymmetry and the distance to M87. However, RP stripping from the M87 halo may not be the sole explanation for the range of \hi\, morphologies, but some of the galaxies may have gone through the M49 halo, causing their \hi\, morphology to be disturbed.

The two-dimensional distribution of the detections suggests that they are mostly located in the filamentary structure defined by the X-ray distribution in Fig. \ref{fig:rosat-hi}. Their systemic velocities listed in Table \ref{tb:maintable} also suggests that all the detections, except NGC 4470, belong to the Virgo cluster. However, considering their Tully-Fisher distances listed in column (6) of Table \ref{tb:maintable}, it is clear that NGC 4451, NGC 4411B, NGC 4416, NGC 4466 and VCC 0952 are not in the filament. They appear to be background galaxies, although their systemic velocities may suggest otherwise. Apart from these galaxies, all the other detections may well belong to the filament.

\subsection{Gas content}\label{sec:gas-content}
To evaluate the relative gas content of the detections, we determine their distance-independent \hi\, deficiency parameter; this is defined as  \citep[][hereafter Sol02]{Solanes2002}
\begin{equation}\label{eq:def}
{\rm def_\HI} = \langle \log{\overline{\Sigma}_\HI (D,T)}\rangle - \log{\overline{\Sigma}_\HI},
\end{equation}
where the mean {\it hybrid} \hi\, surface density is obtained from the ratio of the integrated \hi\, flux density $S_{\HI}$ of the galaxy in $\rm Jy\,km\,s^{-1}$ and its apparent optical diameter $a$ in arcminutes \citep{Solanes1996}:
\begin{equation}\label{eq:hyb-surfdens}
\overline{\Sigma}_\HI = S_{\HI} / a^2.
\end{equation}
The term $\langle\log{\overline{\Sigma}_{\HI}(T)}\rangle$ is the base-10 logarithm of the expected {\it hybrid} \HI\, column density value for a field galaxy of the same diameter $D$ and morphological type $T$. The higher the deficiency parameter, the more \hi-poor the galaxy is relative to the mean of a comparison sample of field galaxies. Column (10) of Table \ref{tb:maintable} lists the values of ${\rm def_\HI}$ for the detected galaxies. Among the 13 detected galaxies with clear morphological type and having optical data, only 3 (namely UGC 7590, VCC 1142 \& 0952) are not \hi-deficient.

To put these quantities into context, we compare the \hi\, deficiency of the detections to the VLA Imaging survey of Virgo galaxies in Atomic gas \citep[VIVA,][]{Chung2009} sample, and to the sample of ``Local Orphan Galaxies'' \citep[LOG,][]{Karachentsev2011}. The VIVA sample consists of 53 late type galaxies, mostly selected to cover a range of star formation properties based on \citet{Koopmann2004a}'s classification, and both in high and low density environments. The galaxies span the angular distances of $1\dg$ to $12\dg$ from M87 (out to $\sim1.6$ Abell radii of the cluster). \citet{Chung2009} provide the \hi\, masses and distance-independent deficiency parameters for all the galaxies except for two (IC 3418 and VCC 2062) where the \hi\, deficiency is not derived.

The LOG catalog is made of 517 local isolated galaxies selected from HyperLEDA\footnote{http://leda.univ-lyon1.fr} and NED\footnote{http://ned.ipac.caltech.edu}. Of these, 278 are classified as irregular galaxies, known to be gas-rich. Of the remaining 239 spirals, only 186 have their optical sizes listed in HyperLEDA. We hereafter refer to these as the LOG sample for our comparisons. The optical B-band diameters (at the 25th magnitude) of the LOG galaxies, combined with their \hi\, line flux, distance and morphological type provided in \citet{Karachentsev2011}, were used to evaluate their \hi\, deficiency parameters. The \hi\, masses were derived using
\begin{equation}
M_\HI = 2.36\e{5}\, d^2\, S_\HI
\end{equation}
where $d$ is the distance of the galaxy in Mpc and $S_\HI$ is the flux in Equation \ref{eq:hyb-surfdens}.

\begin{figure}
		\hspace{-20pt}
		\includegraphics[width=1.25\columnwidth]{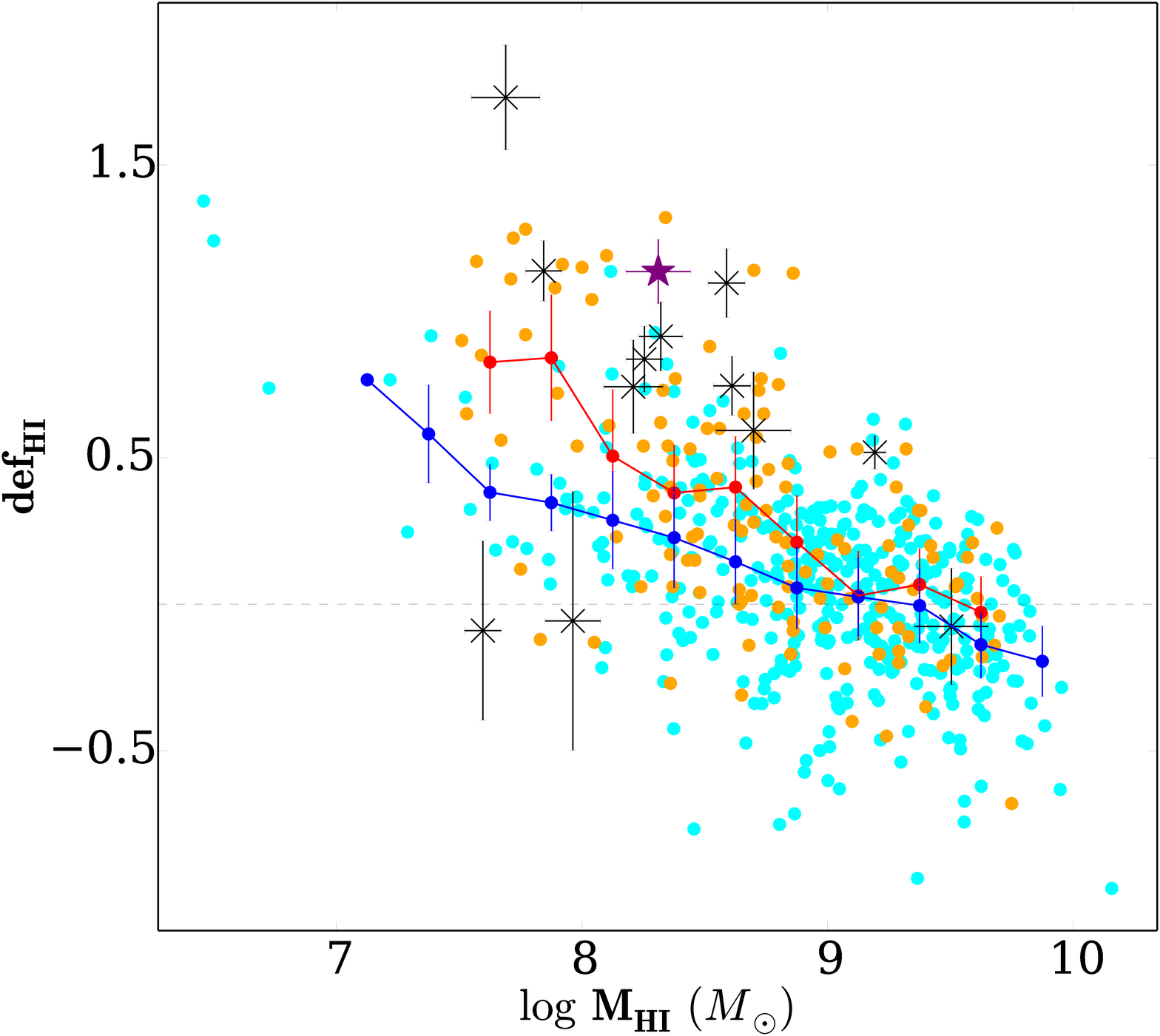}
		\caption{\hi\, deficiency parameter as a function of the \hi\, mass for the different samples. The lines represent the averaged bins. The {\it cyan dots} and {\it blue line} are the LOG sample, and the {\it orange dots} and the {\it red line} show the VIVA sample. The {\it black symbols} represent the detections in this work (the {\it crosses} are the 5 background galaxies discussed in section \ref{sec:hi-morph}, and the {\it diamonds} are believed to belong to the filament), and the {\it purple star} denotes the one-sided \hi-tail'ed spiral NGC 4424.}\label{fig:defhivsmhi}
\end{figure}

In Fig. \ref{fig:defhivsmhi} we present the \hi\, deficiency parameter as a function of \hi\, mass for our detection sample over-plotted on the VIVA and LOG samples. The \hi\, deficiency presents little variation with the \hi\, mass for the field galaxies (the LOG sample); it slightly decreases as one moves to higher \hi\, masses, but remains close to $\rm def_\HI = 0$ throughout the probed \hi\, mass range. However, for the galaxies in the vicinity of the Virgo Cluster (the VIVA and the detection samples), the deficiency parameter decreases with increasing \hi\, mass.

Out of the three non-deficient galaxies, only UGC 7590 has a high \hi\, mass. The other two detections, VCC 1142 and VCC 0952, have \hi\, masses $\lesssim 10^8\,\Mo$. Figure \ref{fig:rosat-hi} shows that these galaxies are not confined in a same region, VCC 1142 being near the virial radius and VCC 0952 well within. However, their Tully-Fisher distances distribution (see Table \ref{tb:maintable}) suggest that VCC 0952 is a background galaxy, making it normal for the galaxy to be gas-rich. The relatively high \hi\, mass of UGC 7590 may explain the non-deficiency of the galaxy, as it falls in the lower-right corner of the plot in Figure \ref{fig:defhivsmhi}. The abundance of \hi\, in VCC 1142 is a surprise, given the location of the galaxy in the cluster (Figure \ref{fig:rosat-hi}) and its low \hi\, mass. Its non-deficiency suggests that, unlike most of the detections, the galaxy is experiencing no \hi\, depletion, but may be accreting gas from the IGM.

The Virgo galaxies at the low-mass end of the sample present much higher deficiencies, with respect to the field galaxies, than galaxies at the high-mass end. If the samples are representative of the galaxies inside the Virgo cluster, this trend may be an indication that most of the cluster's low \hi\, mass galaxies are \hi-deficient, suggesting that the cluster environment of Virgo is more severe on low \hi\, mass galaxies. Most of the galaxies in our detection sample present deficiencies lower than the average of the VIVA sample.
If they were actual members of the filament connecting the M87 group to the M49 group, one would expect a depletion of their \hi\, content due to interactions with the hot X-ray gas, i.e, they should present higher \hi\, deficiencies with respect to the other cluster members.

In fact, the galaxies in the VIVA sample were selected to represent late type Virgo spirals in morphological type, systemic velocity, subcluster membership, \hi\, mass, and deficiency. Given that most of our detections (including those believed to belong to the filament) lie below the average VIVA galaxies in Fig. \ref{fig:defhivsmhi}, we conclude that the \hi\, depletion of the galaxies in the filament is not more pronounced than those in the M87 halo.

\subsection{Case of NGC 4424}
Several authors argued that the tail seen in NGC 4424 is a mixed effect of RP and tidal interaction \citep{Kenney1996, Cortes2006, Chung2007, Chung2009}. The arguments for tidal interaction are mainly based on the optical morphology of the galaxy, classified as a {\it chaotic} early spiral in the Uppsala General Catalogue \citep[UGC,][]{Nilson1973}, a peculiar Sa by \citet{Binggeli1985}, and an uncertain barred Sa by \citet{DeVaucouleurs1991}. The $\rm H\alpha$ morphology of the galaxy has been described as {\it truncated/compact} by \citet{Koopmann2004a} and optical observations reveal that it has a heavily disturbed stellar disk, and shows a strong $\rm H\alpha$ emission confined to its central kpc \citep{Kenney1996, Cortes2006}.

However, the present observations reveal an \hi\, tail of NGC 4424 three times longer than previously thought. This fact questions the hypothesis of tidal interaction as the main cause of the tail, given that the optical morphology does not leave room for such a long tail since there is no stellar counterpart to the tail, such as one would expect from a tidal encounter. Moreover, tidal interactions usually create two tails \citep[e.g.,][]{Toomre1972, Yun1994}, while we only detect one long tail here.

The ``dark" galaxy VIRGOHI 21 was observed in the Virgo cluster to have a similar \hi\, tail; it's a starless \hi\, cloud, apparently connected by a faint \hi\, bridge to NGC 4254, a spiral galaxy located $\sim20'$ away \citep{Minchin2007}. Later numerical simulations by \citet{Duc2008} showed that the long tail of the cloud might have resulted from a high-velocity collision, probably with the parent galaxy NGC 4254. The simulations predict the formation of a short counter-tidal tail that falls back onto the parent galaxy, giving a single-tailed morphology to the structure. Although the authors identified the galaxy NGC 4192 as a possible interacting companion for VIRGOHI 21, they also point that the high velocity of the galaxy could move it relatively far for the \hi\, tail. This makes it hard to identify the real interactor, and any massive galaxy in a certain radius could be a potential candidate. This type of process might explain the single tail of NGC 4424, and the galaxy could have collided with any massive galaxy in the Virgo cluster.

\citet{Chung2007} compared the RP to the restoring force in \HI-tail'ed Virgo galaxies using the simulations of \citet{Vollmer2001} and found that for NGC 4424, RP due to the observed hot X-ray emitting gas could exceed the restoring force, meaning that RP is likely at the origin of the \hi\, tail.

Several observations and simulations have shown that RP may be effective in removing gas from cluster galaxies out to $1-2$ virial radii \citep[e.g.][]{Kenney2004a, Crowl2005, Tonnesen2007, Bahe2013}. In Fig. \ref{fig:rosat-hi} we show that NGC 4424 is well within the virial radius of the cluster ($\sim0.8\,r_{\rm vir}$) and therefore may be subject to RP stripping.  By comparison, the only evidence for tidal interaction in NGC 4424 is its complex optical morphology. We argue that this is not sufficient to explain the origin of the tail. A possible scenario would be that a tidal interaction moved the galaxy's \hi\, to a larger radius, making it easier for RP to `blow away' the gas. This would make the RP stripping more efficient, and could well explain the high mass of the \hi\, tail of $(4.3\pm0.9)\e{7}\,\Mo$, i.e $\sim20\%$ of the galaxy's total \hi\, mass.


\section{Summary and Conclusions}\label{sec:summary}

In this work we used the KAT-7 and WSRT telescopes to observe a region of $\sim 2.5\dg\times1.5\dg$ located in the Virgo Cluster, $\sim3\dg$ away from the centre of the cluster and containing the elliptical M49. The distribution of the hot X-ray gas in the field suggests a filamentary structure joining the M49 subcluster to the Virgo cluster. With a total of $\sim 78$ and 48 hours of observations respectively with the KAT-7 and WSRT telescopes, we reached similar column densities sensitivities with the two telescopes. To detect both the low and high resolution features and benefit from the advantages of both telescopes, we combined the two observations. However, because of the large difference in the point source flux sensitivities in the two datasets, the combination could not be done in the Fourier plane, as is widely practised. We have pioneered a new approach, which consists of combining the data in the image plane after converting the cubes from units of flux density to units of column density. This provided us a dataset highlighting features of both the large and small scale structures, and in which we reach an improved $1\sigma$ sensitivity of $\sim8\e{17}\,\rm cm^{-2}$ over 16.5 \kms.

A total of 14 galaxies and 3 \hi\, clouds with \hi\, masses $M_\HI > 10^7\,\Mo$ were detected, including the one-sided \hi-tailed spiral NGC 4424. 10 out of the 14 galaxies were found to be \hi-deficient. Compared to both a sample of other Virgo spirals and field galaxies, most of the detected galaxies are found to be no more \hi-deficient than typical Virgo galaxies, suggesting that gas stripping processes in the region are not more pronounced than elsewhere in the cluster. In particular, NGC 4424 has an \hi-deficiency typical of the cluster galaxies and exhibits a tail observed to be 60 kpc, three times longer than previous observations with the VLA have revealed \citep{Chung2007}. The morphology of the tail, as well as the asymmetry and the \hi\, deficiency observed in other galaxies in the region suggest that RP is most likely the primary mechanism responsible for the tail. A high velocity collision was also found to be a possible cause of the tail, although the companion remains unidentified.


\section*{Acknowledgments}

We thank all the team of SKA South Africa for allowing us to obtain scientific data during the commissioning phase of KAT-7. 
The work of CC is based upon research supported by the South African Research Chairs Initiative (SARChI) of the Department of Science and Technology (DST), the Square Kilometre Array South Africa (SKA SA) and the National Research Foundation (NRF). The research of AS \& KH have been supported by SARChI, SKA SA fellowships. AS was additionally funded by the National Astronomy and Space Science Programme (NASSP).

\bibliographystyle{mn2e}
\bibliography{./library}

\clearpage

\begin{table}
	\footnotesize
	\begin{center}
	\caption{Properties of detected galaxies.}\label{tb:maintable}
	\begin{tabular}{ l  c c c c c c c c c c c c c c c c c c}
      \hline\hline
      \multicolumn{1}{c}{Object} & R.A & Dec.   &  Type & $D_{25}$ & $i$ & $d$ & $v_{\rm sys}$ & $W_{50}^c$ & $M_\HI$ & $\rm def_{\HI}$ & $d_{\rm M87}$\\
			 
			 &  \multicolumn{2}{c}{(J2000)}			&   & $(')$& (deg) & (Mpc) & (\kms)		& (\kms)	 &  $(10^8\,\Mo)$ &  & (deg) \\
					
      \multicolumn{1}{c}{(1)} & \multicolumn{2}{c}{(2)} & (3) & (4) & (5) & (6) & (7) & (8) & (9) & (10) & (11)\\ 
      \hline
      NGC 4424  & 12 27 11.6 & 09 25 14 & SBa & 3.63 & 62.1 & $15.2\pm1.9^a$ & 433 &   58.6 & $2.0\pm0.4$ & $0.79\pm0.09$ & 3.10\\
      NGC 4451  & 12 28 40.5 & 09 15 31 & Sab & 1.48 & 51.2 & $28.7\pm1.0$ & 864 &  255.8 & $5.4\pm0.6$ & $0.14\pm0.17$ & 3.18\\
      NGC 4470  & 12 29 37.8 & 07 49 27 & Sa  & 1.29 & 44.7 & $16.2\pm1.6$ & 2321 &  135.7 & $1.6\pm0.3$ & $0.05\pm0.14$ & 4.58\\
      UGC 7590  & 12 28 18.8 & 08 43 46 & Sbc & 1.35 & 76.6 & $21.9\pm2.9$ & 1112 &  177.2 & $16.5\pm3.5$ & $-0.50\pm0.17$ & 3.71\\
	    NGC 4411A & 12 26 30.0 & 08 52 18 & Sc  & 2.04 & 54.4 & $15.1\pm1.5$ & 1271 & 105.2 & $1.8\pm0.3$  & $0.44\pm0.09$ & 3.68\\
	    NGC 4411B & 12 26 47.2 & 08 53 04 & Sc  & 2.51 & 26.7 & $27.9\pm0.6$ & 1260 & 153.9 & $15.6\pm1.6$  & $0.21\pm0.06$ & 3.64\\
	    UGC 7557  & 12 27 11.1 & 07 15 47 & Sm	& 3.02 & 21.3 & $16.8\pm1.7^b$ & 924  & 245.2 & $4.1\pm0.7$ & $0.59\pm0.08$  & 5.21\\
	    NGC 4445  & 12 28 15.9 & 09 26 10 & Sab & 2.63 & 90.0 & $19.1\pm0.9$ & 418 & 171.8 & $0.4\pm0.1$ & $1.36\pm0.14$ & 3.02 \\
	    VCC 1142  & 12 28 55.5 & 08 49 01 & dE  & 0.27 & 53.4 & $20.1\pm2.0^c$ & 1334 & 52.0 & $0.4\pm0.1$ & $-0.33\pm0.33$ & 3.60 \\
	    NGC 4416  & 12 26 46.7 & 07 55 08 & Sc  & 1.70 & 24.0 & $36.6\pm4.0$ & 1381 & 229.2 & $3.3\pm0.6$ & $0.78\pm0.09$ & 4.58 \\
	    NGC 4466  & 12 29 30.6 & 07 41 47 & Sab & 1.32 & 74.9 & $27.7\pm3.3$ & 797 & 185.9 & $2.1\pm0.4$ & $0.42\pm0.11$ & 4.71 \\
	    UGC 7596  & 12 28 33.9 & 08 38 23 & Im  & 1.66 & 71.9 & $16.4\pm1.6^d$ & 595 & 59.5 & $0.7\pm0.1$ & $0.82\pm0.09$ & 3.79 \\
	    VCC 0952  & 12 26 55.7 & 09 52 56 & SABc & 0.26 & 54.6 & $24.8\pm4.2$ & 1024 & 100.3 & $0.9\pm0.2$ & $-0.63\pm0.48$ & 2.68 \\
	    AGESVC1 293 & 12 29 59.1 & 08 26 01 & ? & $0.57^e$ & $41.8^e$ & $17.0\pm1.7^e$ & 615 & 87.3 & $0.2\pm0.1$ & -- & 3.96 \\
      Cloud 7c  & 12 30 25.8 & 09 28 01 & \HI\, cloud &  -- & -- & $16.8\pm1.7^b$ &  496 & 74.0 & $0.6\pm0.1$ & -- & 2.93\\
	    M49 Cloud & 12 29 54.4 & 07 57 57 & \HI\, cloud & -- & -- & $16.8\pm1.7^b$ & 476 & 66.0 & $0.7\pm0.1$ & -- & 4.43 \\
	    KW Cloud  & 12 28 34.4 & 09 18 33 & \HI\, cloud & -- & -- & $16.8\pm1.7^b$ & 1270 & 73.2 & $0.7\pm0.1$ & -- & 3.13 \\
      \hline
	\end{tabular}
	\caption{(1) Object name.\\ (2) Optical positions.\\ (3) Morphological type from RC3.\\ (4) Optical diameter at surface brightness $\mu_B=25$ \marc\, from the RC3 catalogue (see notes).\\ (5) Inclination defined as $\cos^2{i} = (q^2 - q_0^2)/(1 - q_0^2)$ where $q$ is the axial ratio of the galaxy and $q_0 = 0.2$ \citep{Aaronson1980}. For NGC 4411 A\&B, the inclination was taken from HyperLEDA\\
(6) Distance of the galaxy from Sol02 (see notes).\\
(7) Systemic velocity.\\ (8) \HI\, profile width, corrected for instrumental effects, turbulence effects, redshift, and inclination, except for the \HI\, clouds where the correction for inclination has not been applied.\\ (9) The \HI\, mass in solar units, assuming that the \HI\, is optically thin.\\ (10) \HI\, deficiency parameter defined in section \ref{sec:gas-content}. \\ (11) Projected angular distance from the elliptical M87 in degrees.\\
 Notes: $^a$ \citealt{Cortes2008}, $^b$ Distance of M87 with a 10\% error, $^c$ HyperLEDA \citep{Makarov2014}, $^d$ \citealt{Haynes2011}, $^e$ \citealt{Taylor2012}.}
	\end{center}
\end{table}

\clearpage

\appendix

\section{Appendix: \hi\, maps and velocity profiles of the detections}
\begin{figure}[]
	\includegraphics[width=0.7\textwidth]{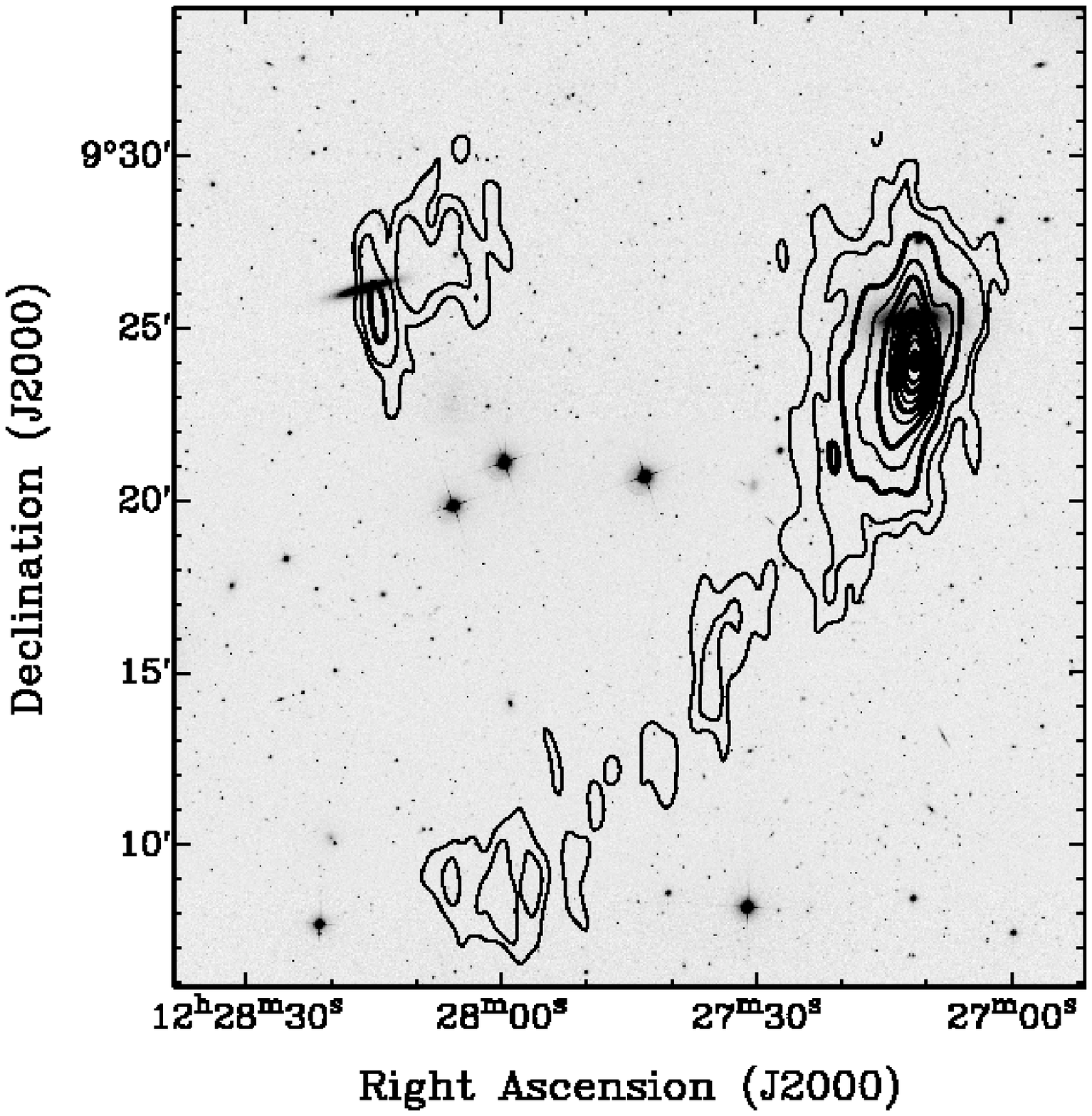}
	\includegraphics[width=0.4\textwidth]{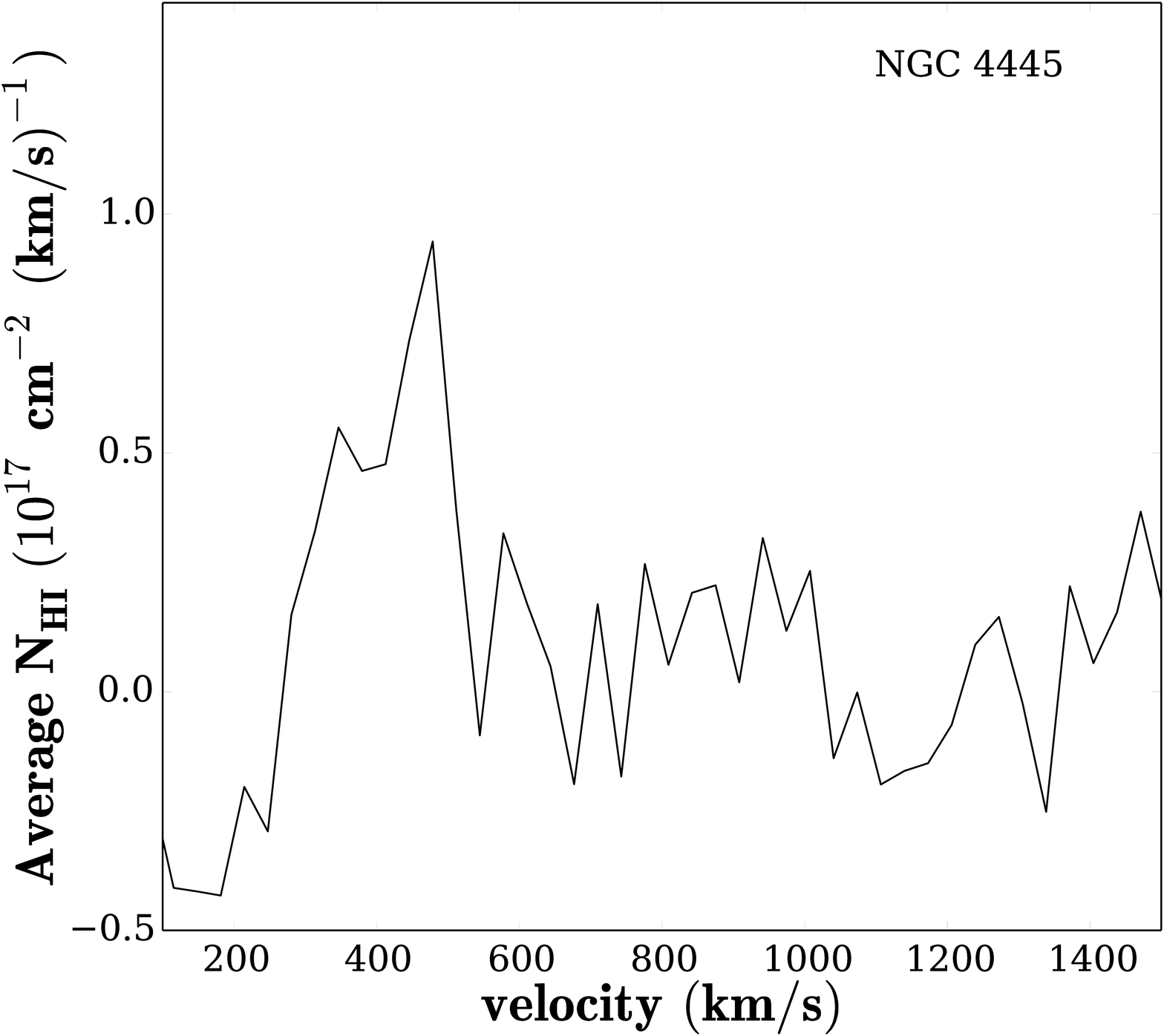}
	\includegraphics[width=0.4\textwidth]{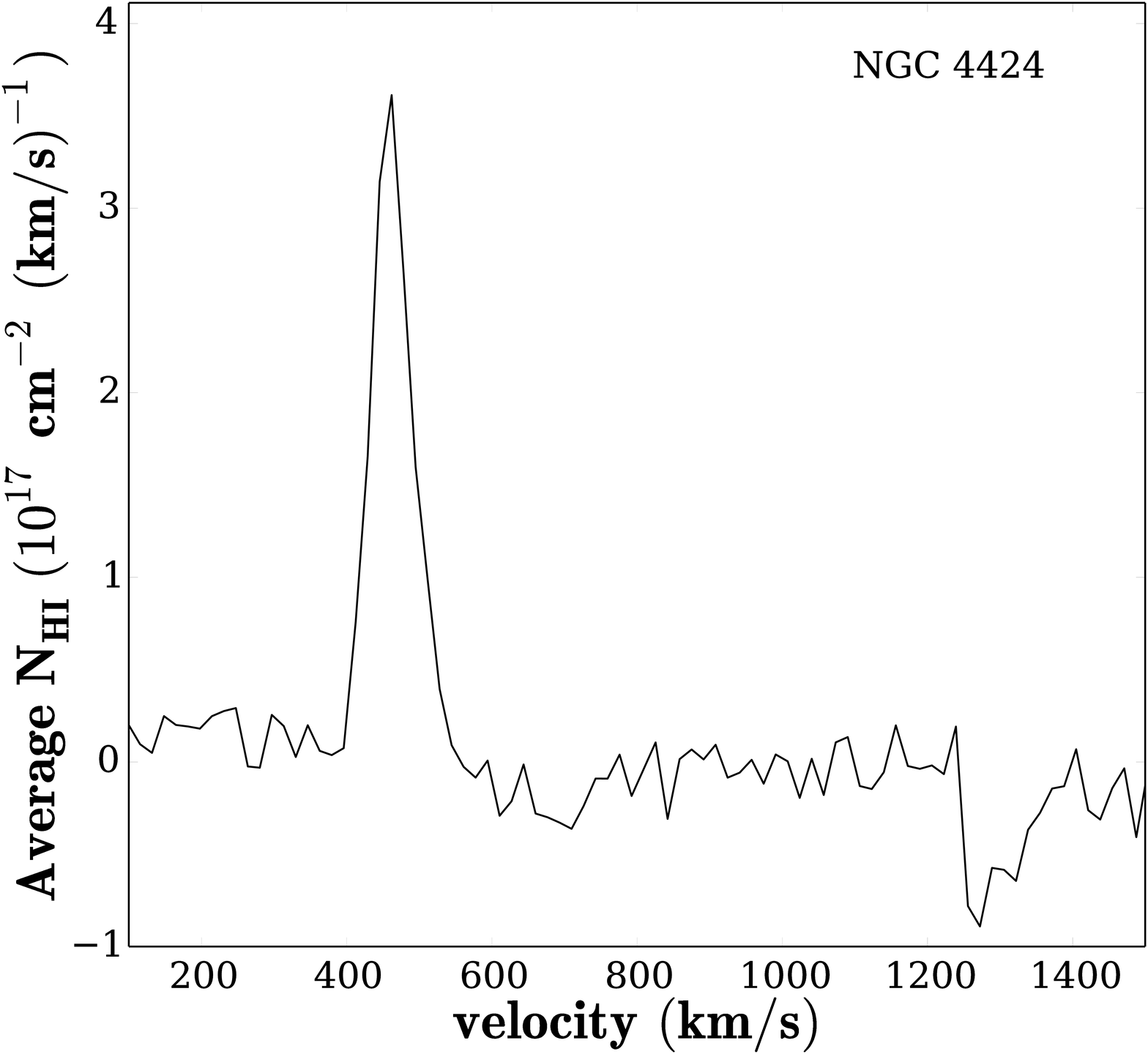}
	\includegraphics[width=0.4\textwidth]{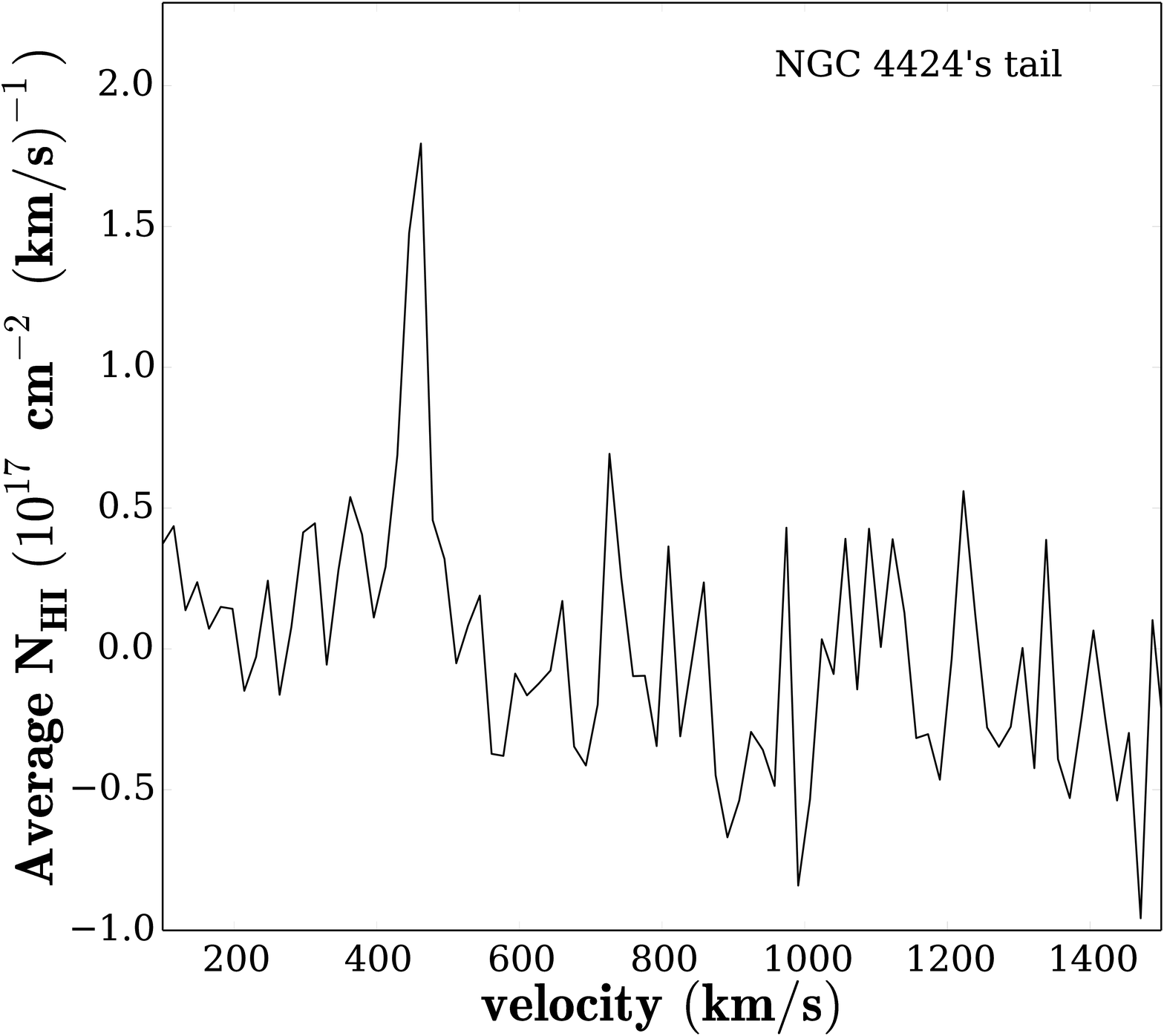}
	\caption{{\it Column density map:} the contour levels are 0.5, 1, 2, 3 ... 10 $\times10^{19}\,\acm$ ({\it left} contours are NGC 4445 and {\it right} are NGC 4424).
	{\it \hi\, profiles: upper right:} NGC 4445 ($v_{\rm sys} = 418$ \kms), {\it lower left:} NGC 4424 ($v_{\rm sys} = 433$ \kms), {\it lower right}: NGC 4424's tail ($v_{\rm sys} = 453$ \kms).}\label{app:n4424}
\end{figure}

\begin{figure}
	\centering
	\includegraphics[width=0.75\textwidth]{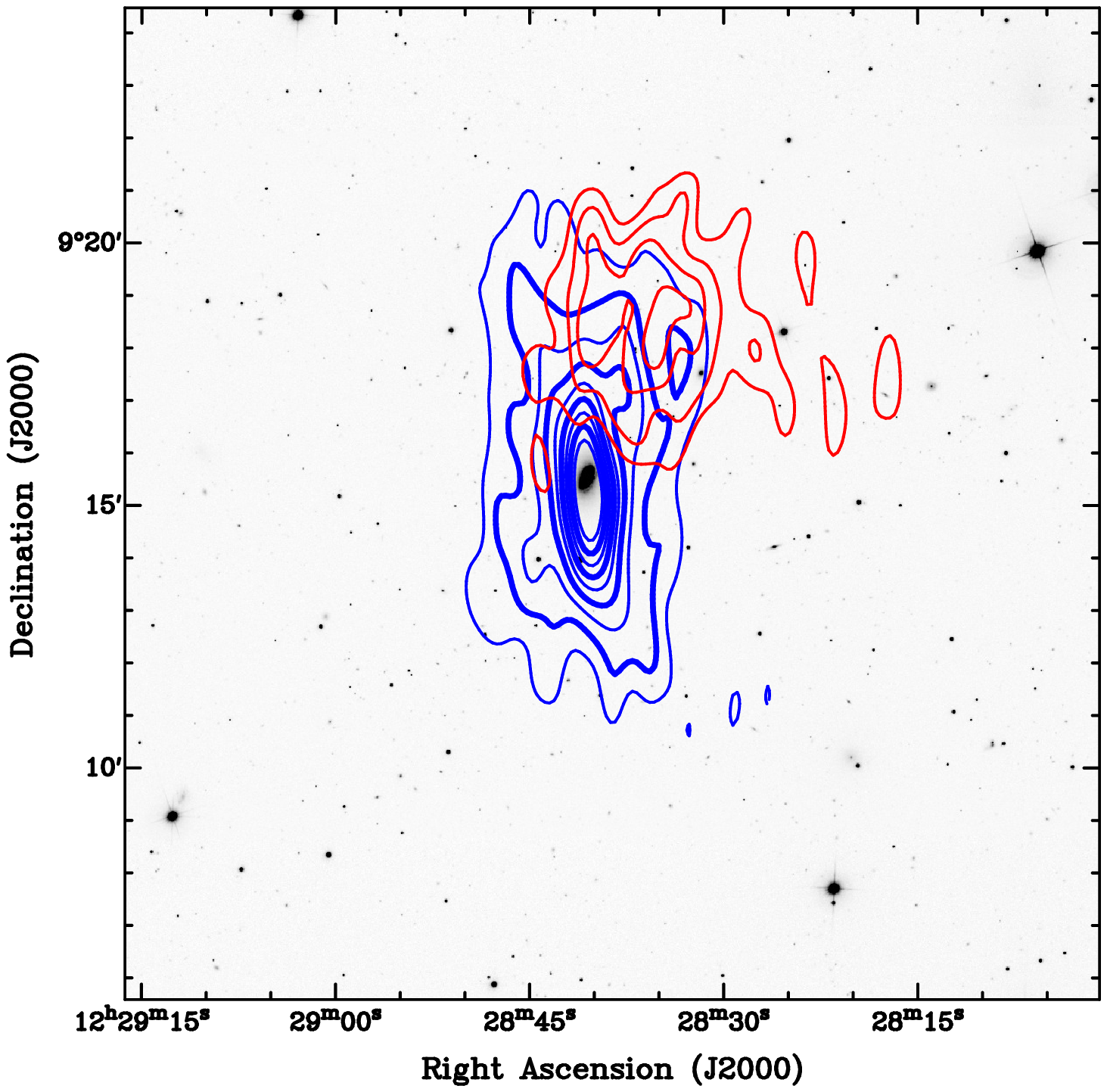}
	\includegraphics[width=0.47\textwidth]{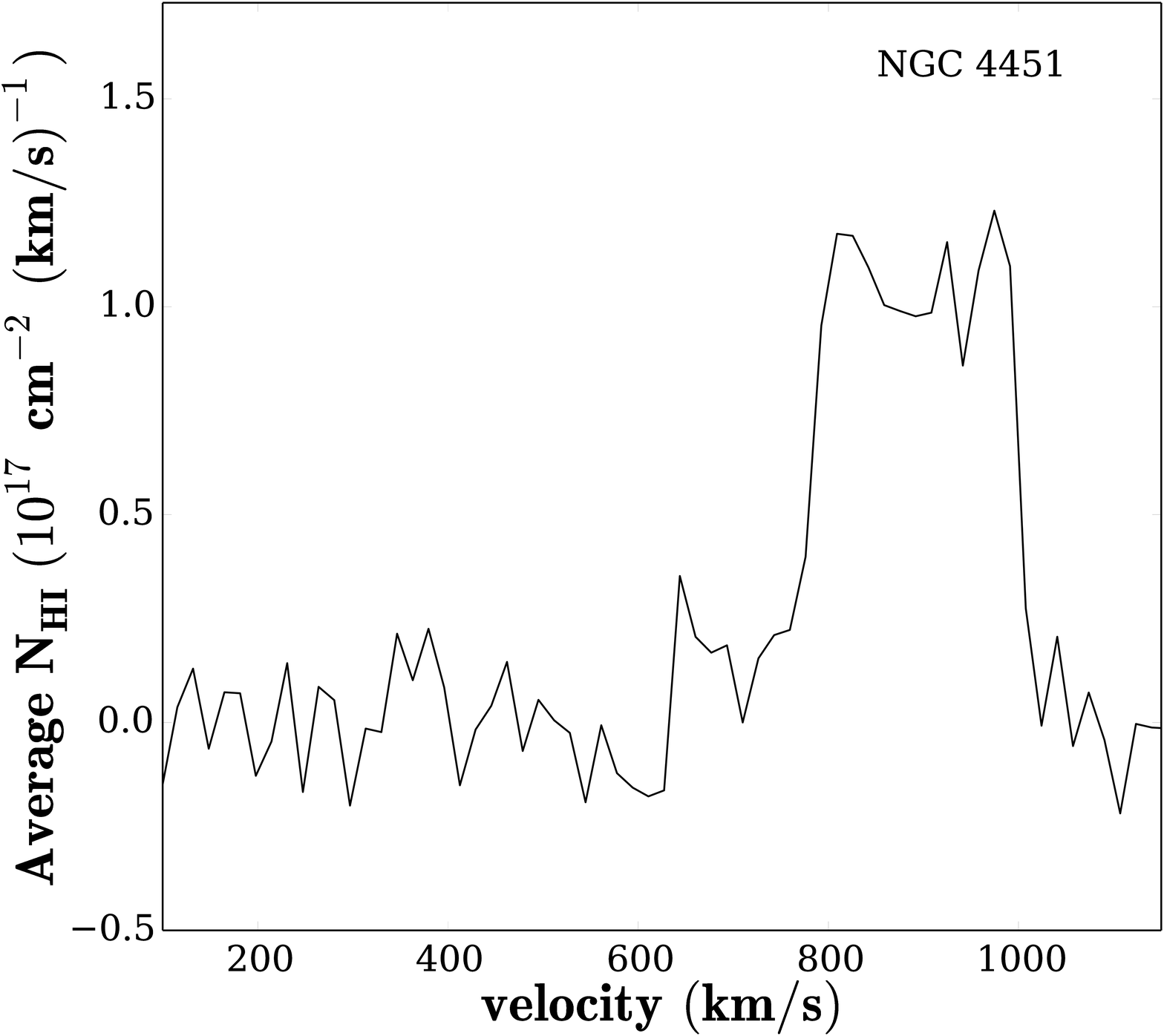}
	\includegraphics[width=0.47\textwidth]{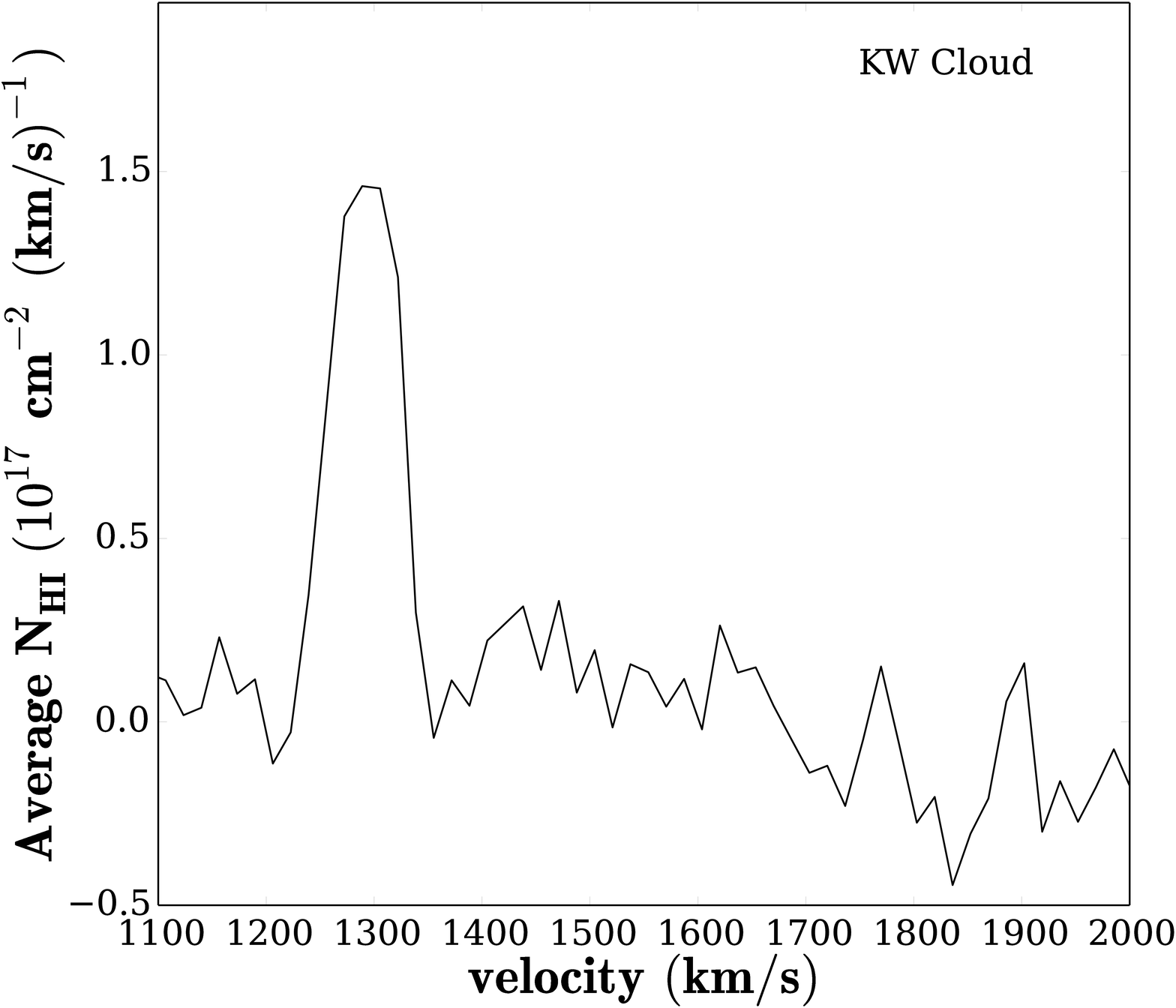}
	\caption{{\it Column density map:} the contour levels of NGC 4451 ({\it blue contours}) are 1, 2, ..., 9 $\times10^{19}\,\acm$ and those of the KW gas cloud ({\it red contours}) are 1, 1.5, 2, 2.5 $\times10^{19}\,\acm$.
	{\it \hi\, profiles: left:} NGC 4451 ($v_{\rm sys} = 864$ \kms), {\it right:} KW cloud ($v_{\rm sys} = 1270$	\kms).}
	\label{app:n4451}
\end{figure}

\begin{figure}
	\centering
	\includegraphics[width=0.75\textwidth]{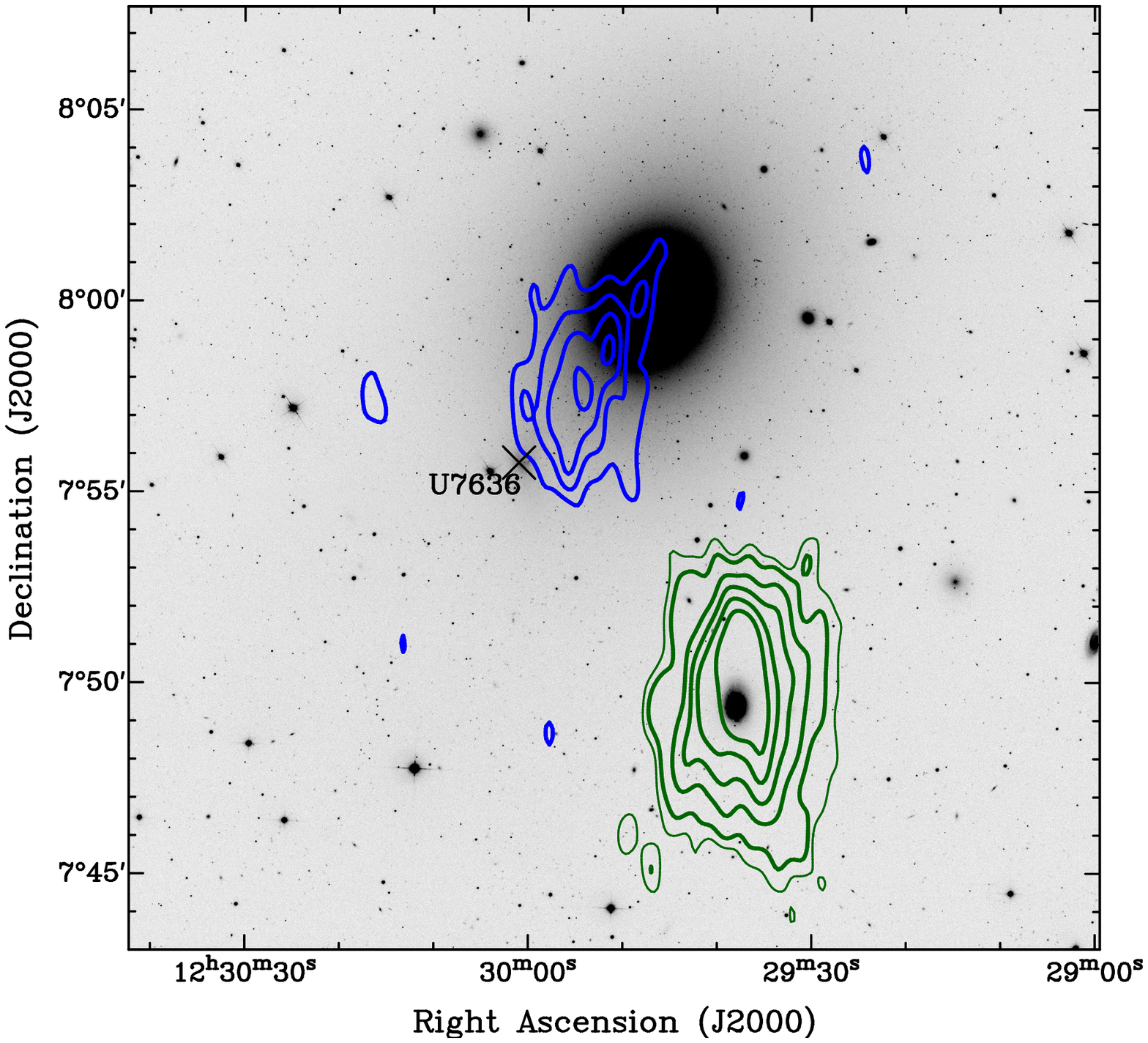}
	\includegraphics[width=0.47\textwidth]{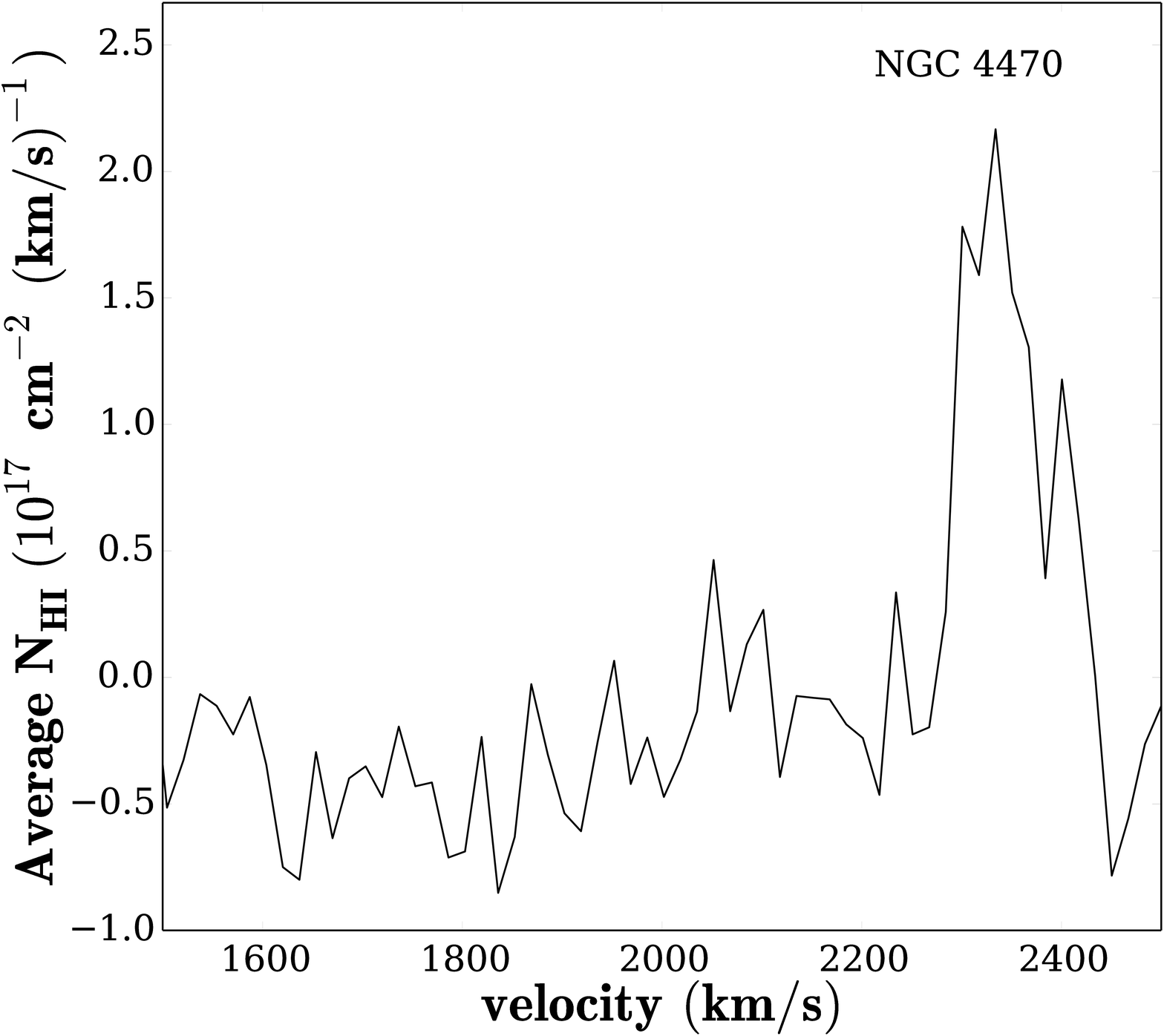}
	\includegraphics[width=0.47\textwidth]{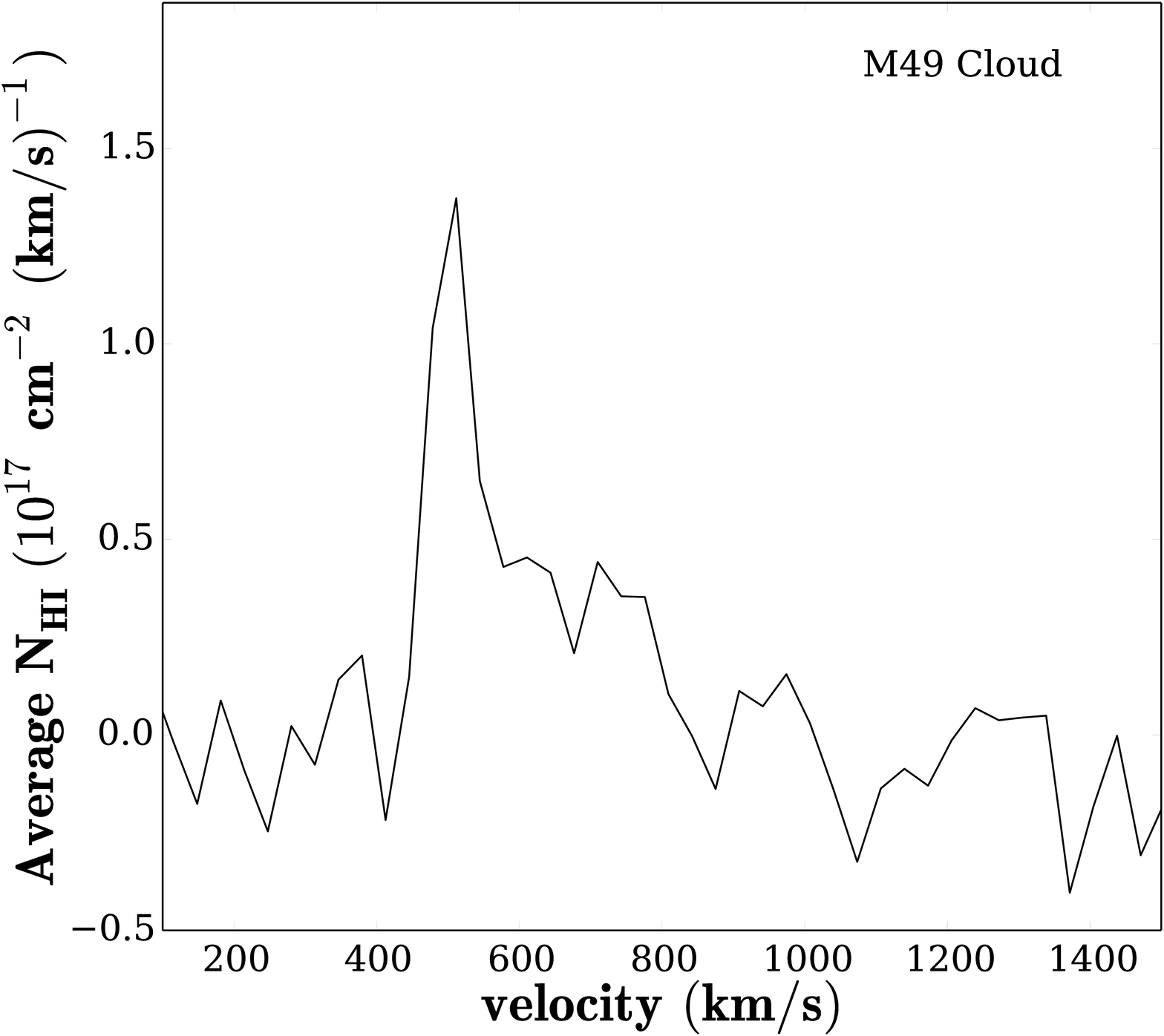}
	\caption{{\it Column density map:} the contour levels of NGC 4470 ({\it green contours}) are 0.5, 1, 2, ..., 5 $\times10^{19}\,\acm$ and those of the M49 gas cloud ({\it blue contours}) are 1, 1.5, 2, 2.5 $\times10^{19}\,\acm$. The dwarf galaxy UGC 7636 ({\it aka} VCC 1249) is denoted by a cross south-east of the cloud.  {\it \hi\, profiles: left:} NGC 4470 ($v_{\rm sys} = 2321$ \kms), {\it right}: M49 cloud ($v_{\rm sys} = 476$ \kms).}
	\label{app:n4470}
\end{figure}

\begin{figure}
	\centering
	\includegraphics[width=0.95\textwidth]{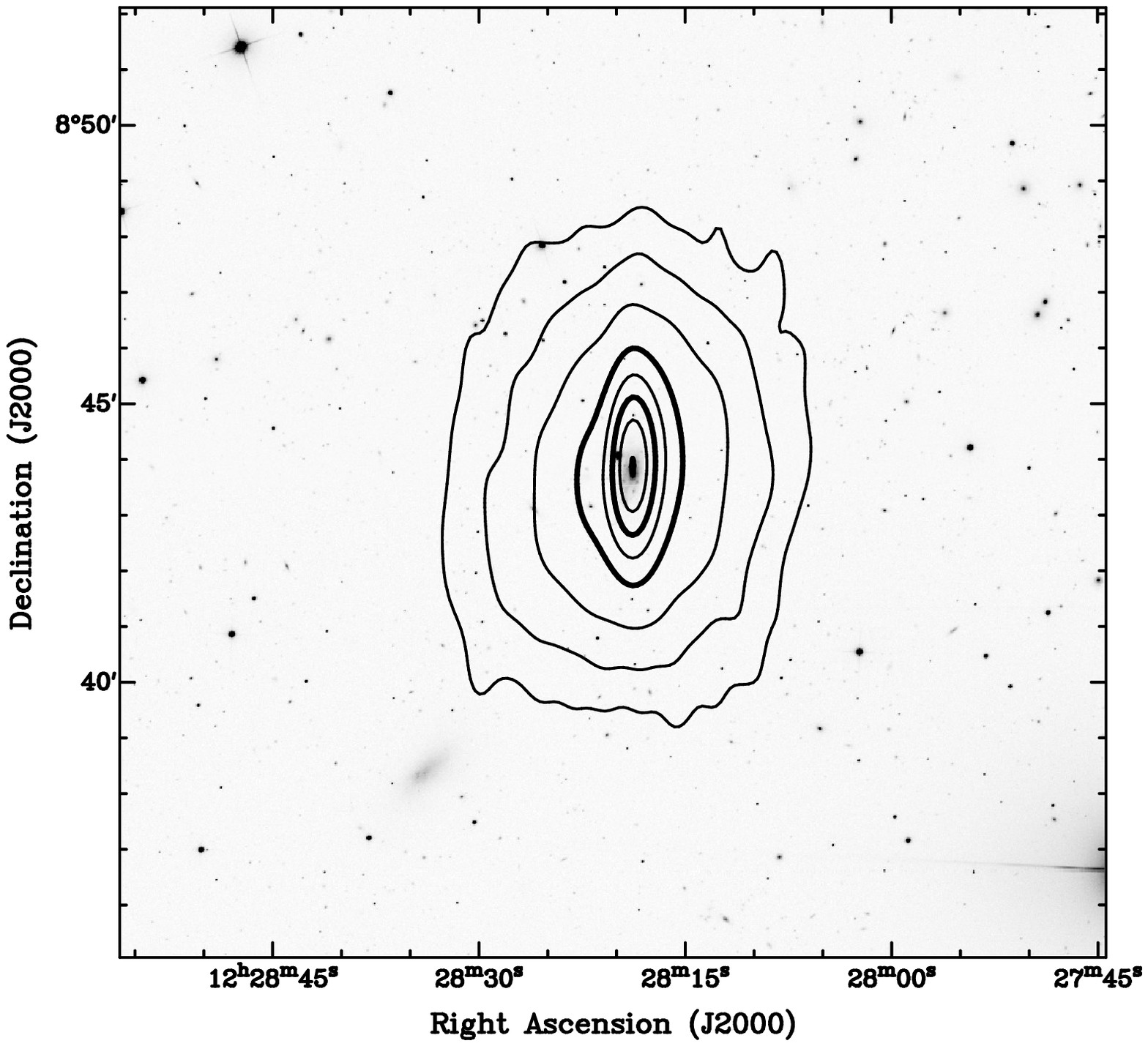}
	\includegraphics[width=0.6\textwidth]{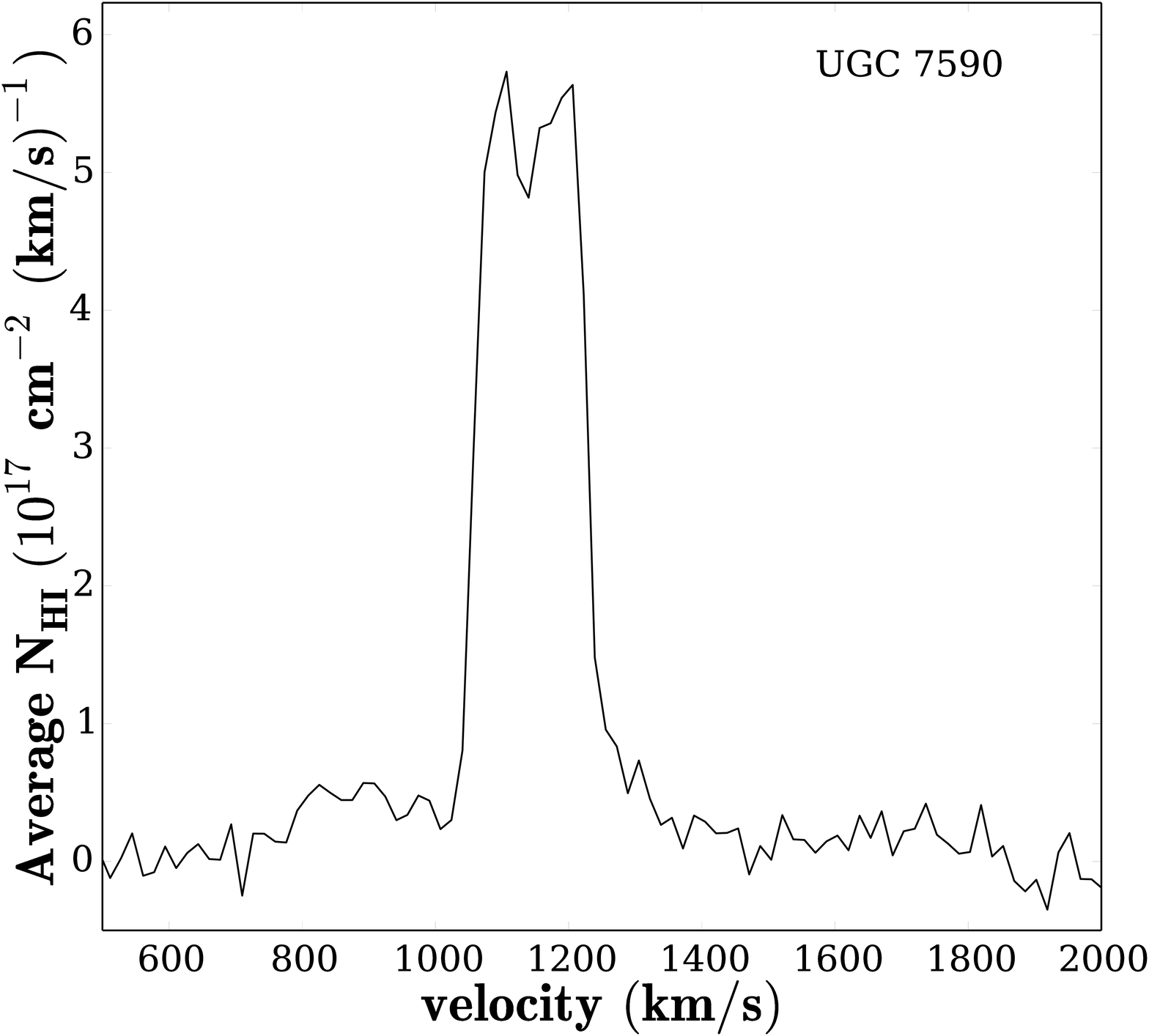}
	\caption{Column density map and \hi\, profile of UGC 7590. Contours levels are 2, 5, 10, 20, 30, ... 60 $\times10^{19}\,\acm$. $v_{\rm sys} = 1112$ \kms.}\label{app:ugc5790}
\end{figure}

\begin{figure}
	\centering
	\includegraphics[width=0.9\textwidth]{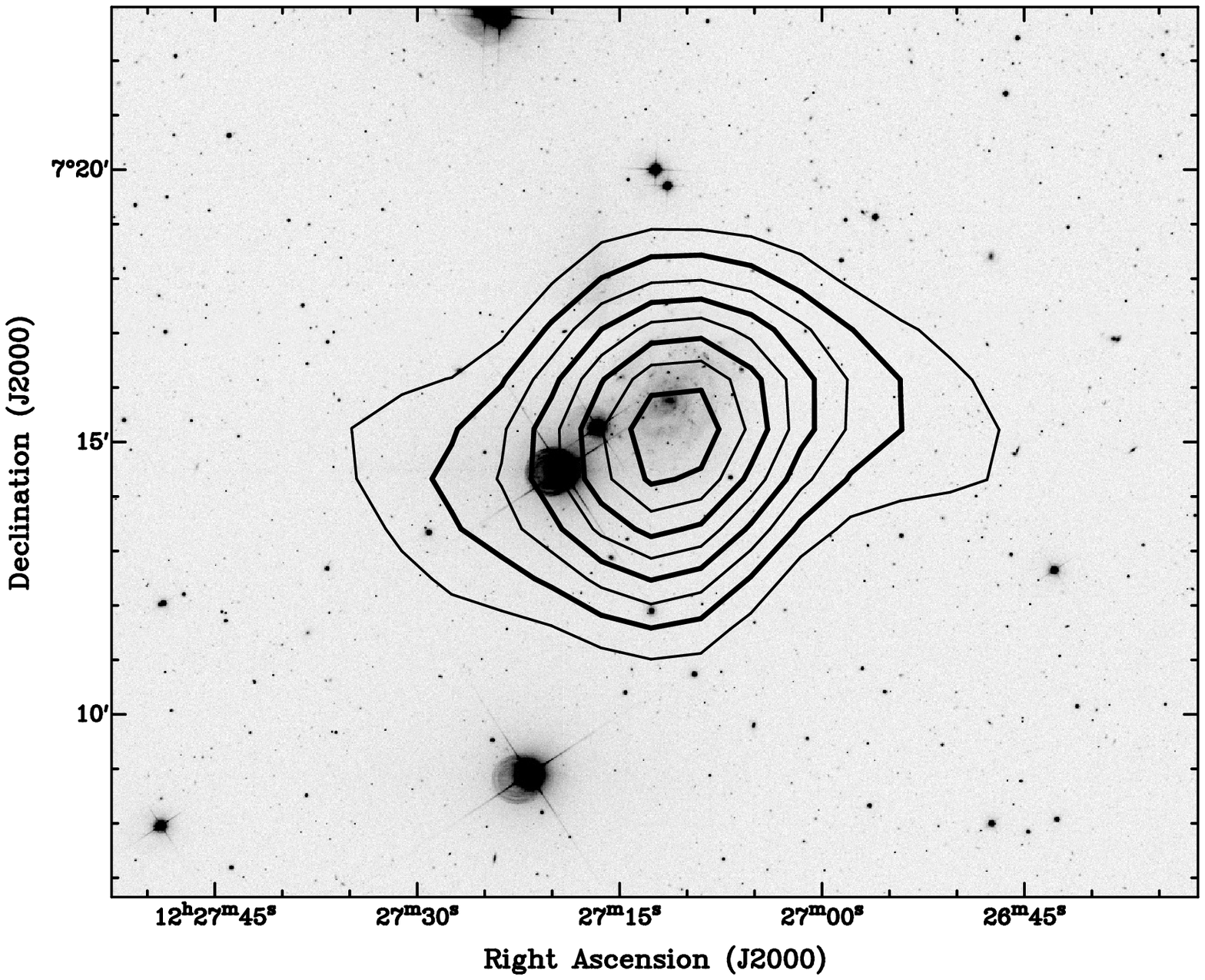}
	\includegraphics[width=0.6\textwidth]{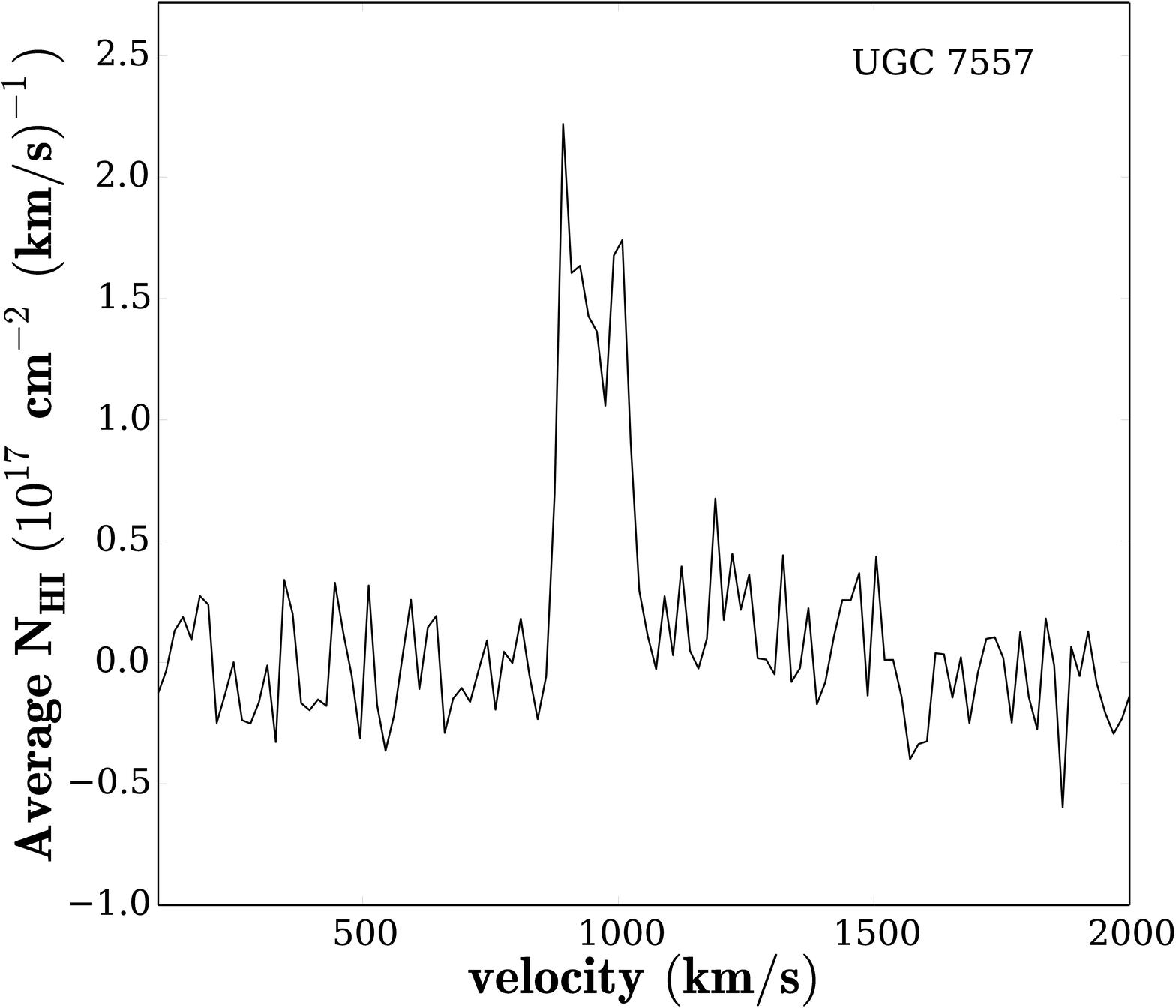}
	\caption{Column density map and \hi\, profile of UGC 7557. Contours: 1, 2, ..., 8 $\times10^{19}\,\acm$. $v_{\rm sys} = 924$ \kms.}\label{app:7557}
\end{figure}

\begin{figure}
	\centering
	\includegraphics[width=0.7\textwidth]{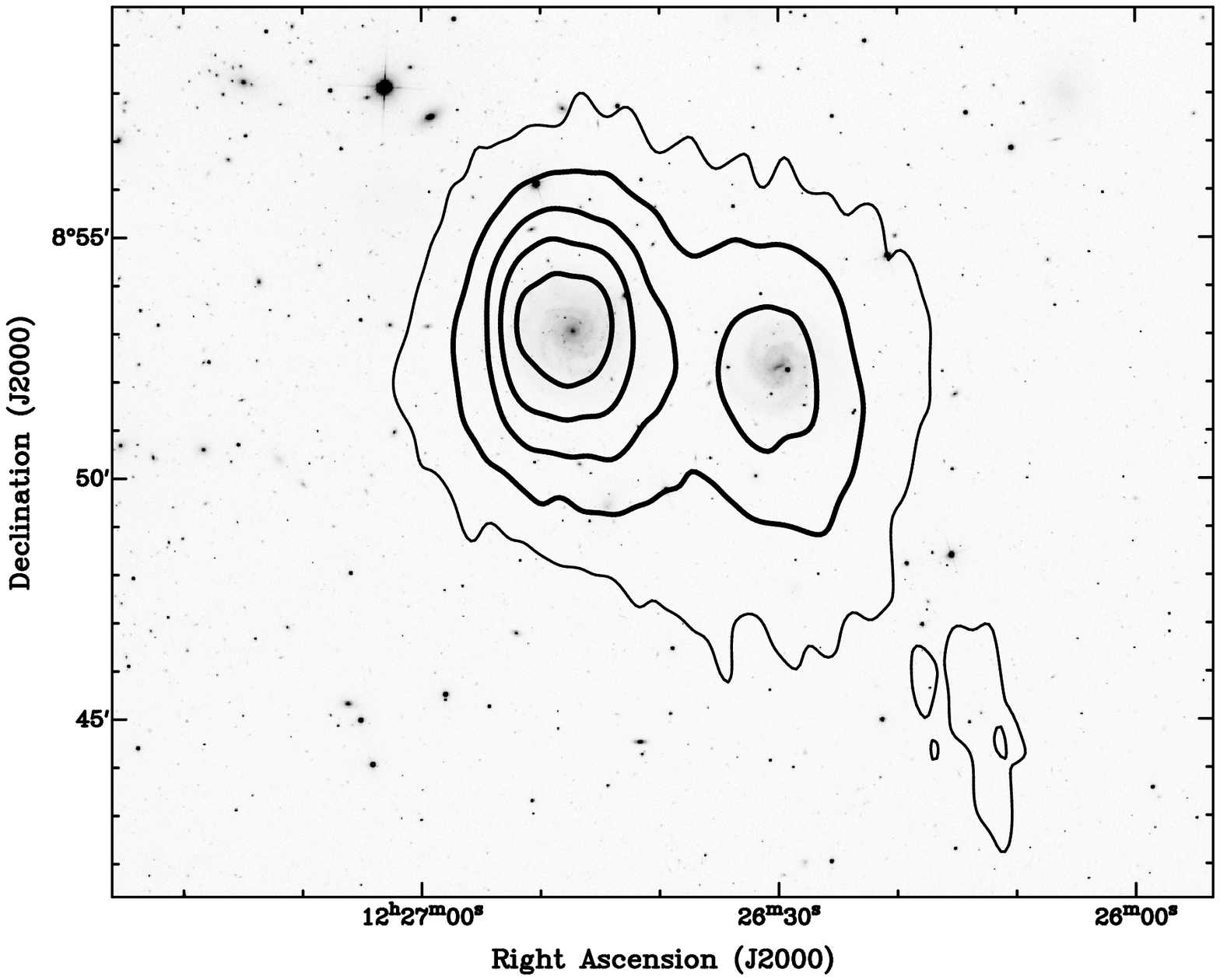}
	\includegraphics[width=0.47\textwidth]{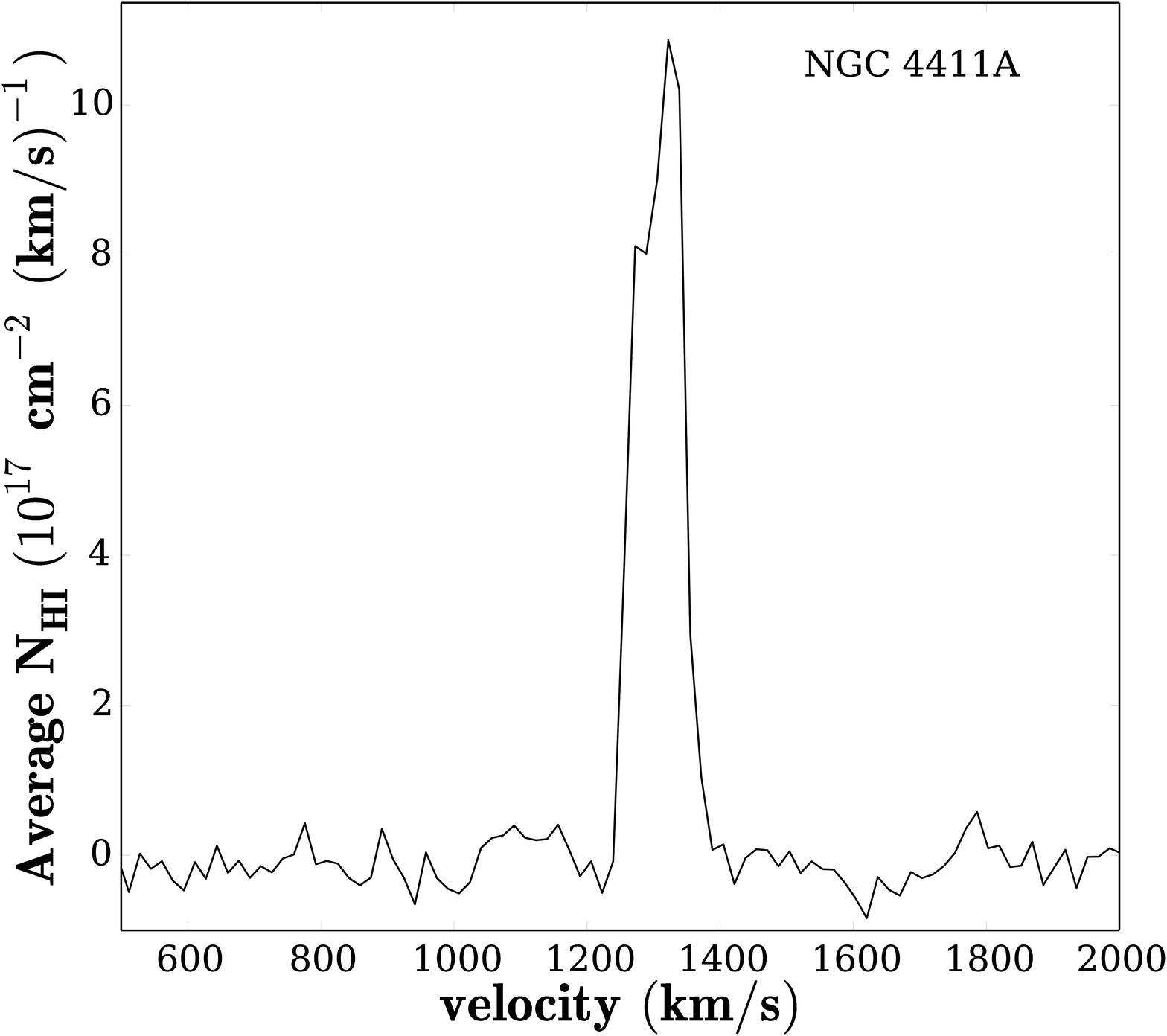}
	\includegraphics[width=0.47\textwidth]{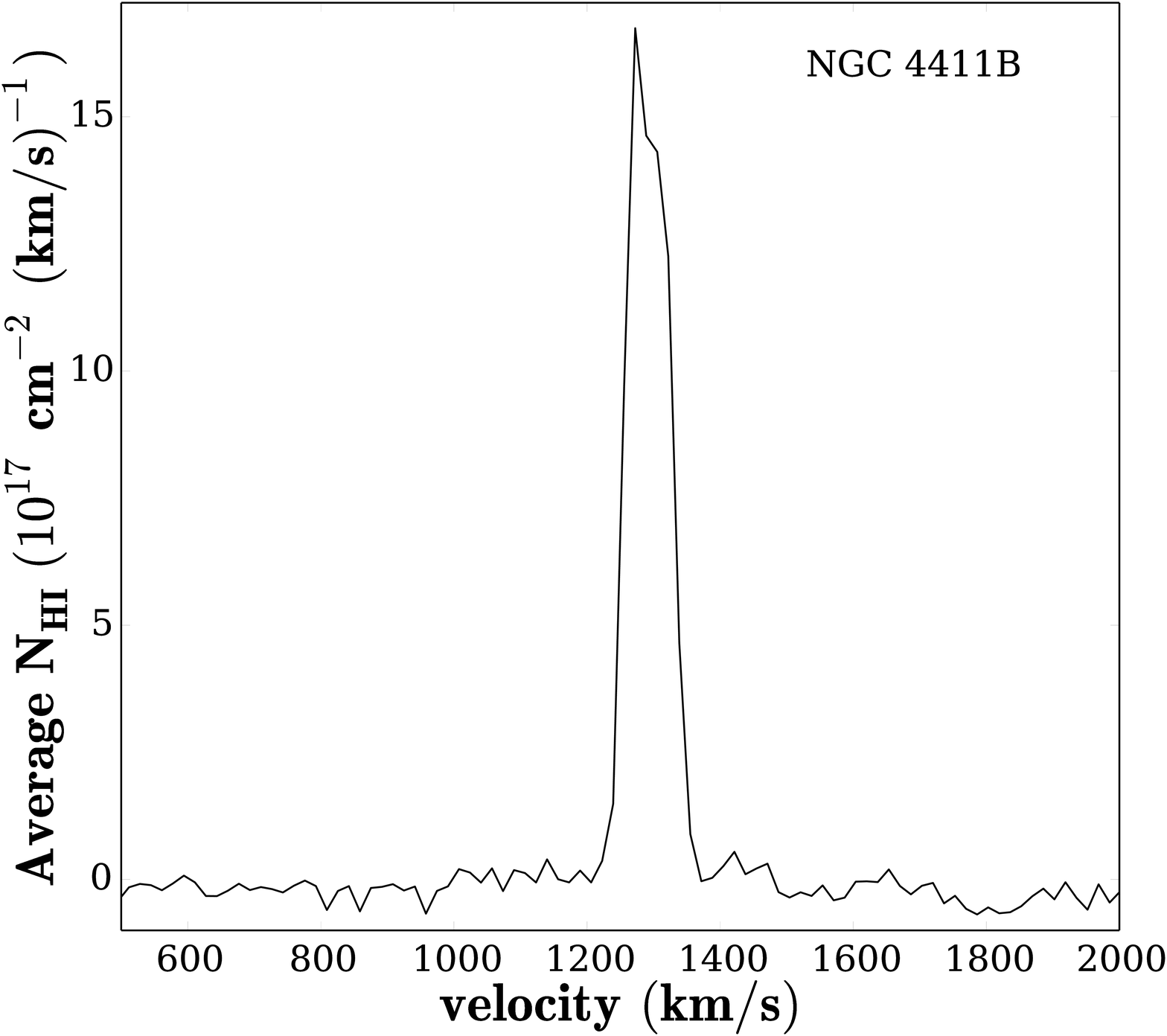}
	\caption{{\it Column density map:} the contour levels are 1, 5, 10, 15, 20 $\times10^{19}\,\acm$ {\it left} contours are NGC 4411B and {\it right} are NGC 4411A). {\it \hi\, profiles: left:} NGC 4411B ($v_{\rm sys} = 1271$ \kms), NGC 4411A ($v_{\rm sys} = 1260$ \kms).}
	\label{app:n4411}
\end{figure}

\begin{figure}
	\centering
	\includegraphics[width=0.9\textwidth]{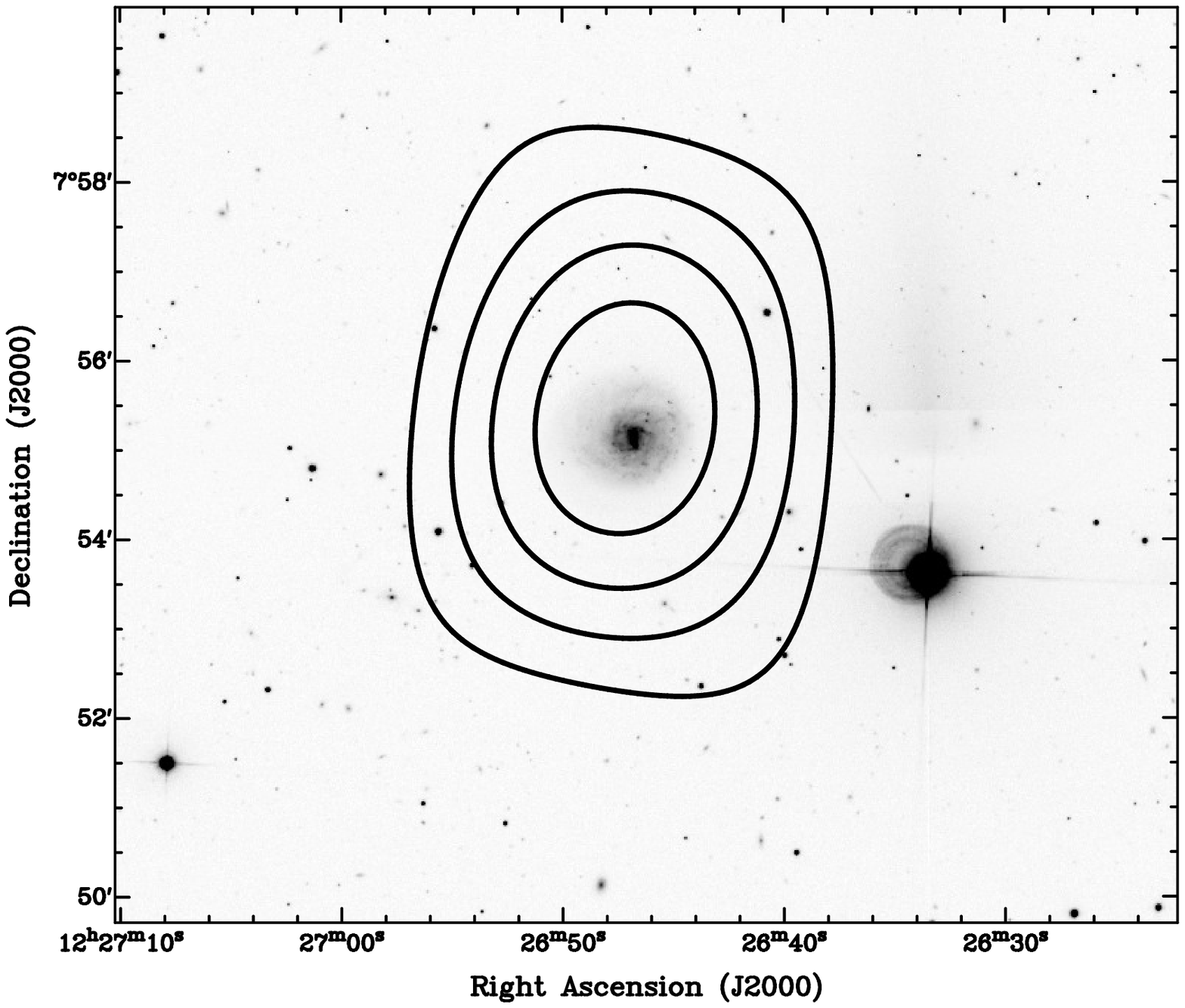}
	\includegraphics[width=0.6\textwidth]{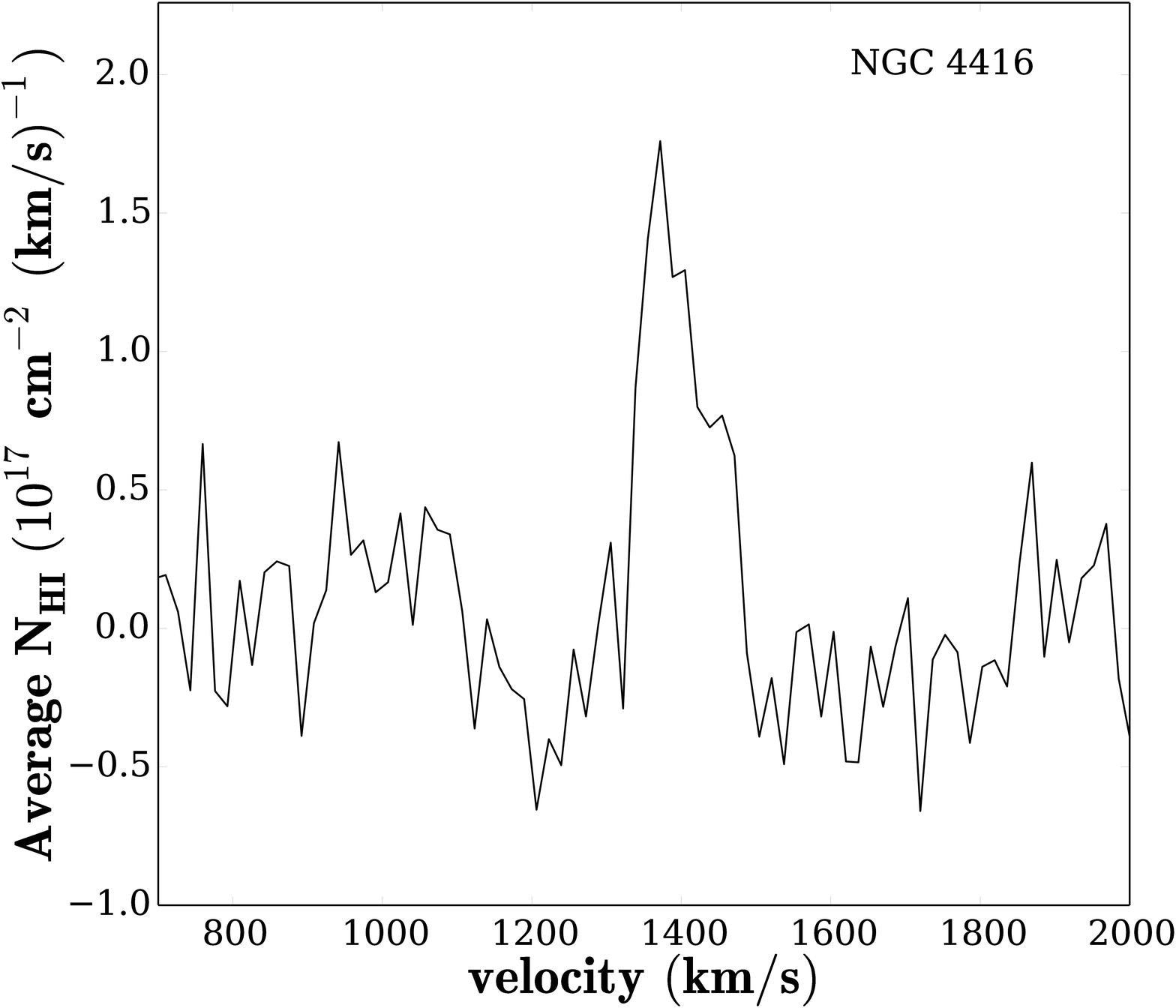}
	\caption{Column density map and \hi\, profile of NGC 4416. Contours: 0.5, 1.0, 1.5, 2.0 $\times10^{19}\,\acm$. $v_{\rm sys} = 1390$ \kms.}\label{app:n4416}
\end{figure}

\begin{figure}
	\centering
	\includegraphics[width=0.85\textwidth]{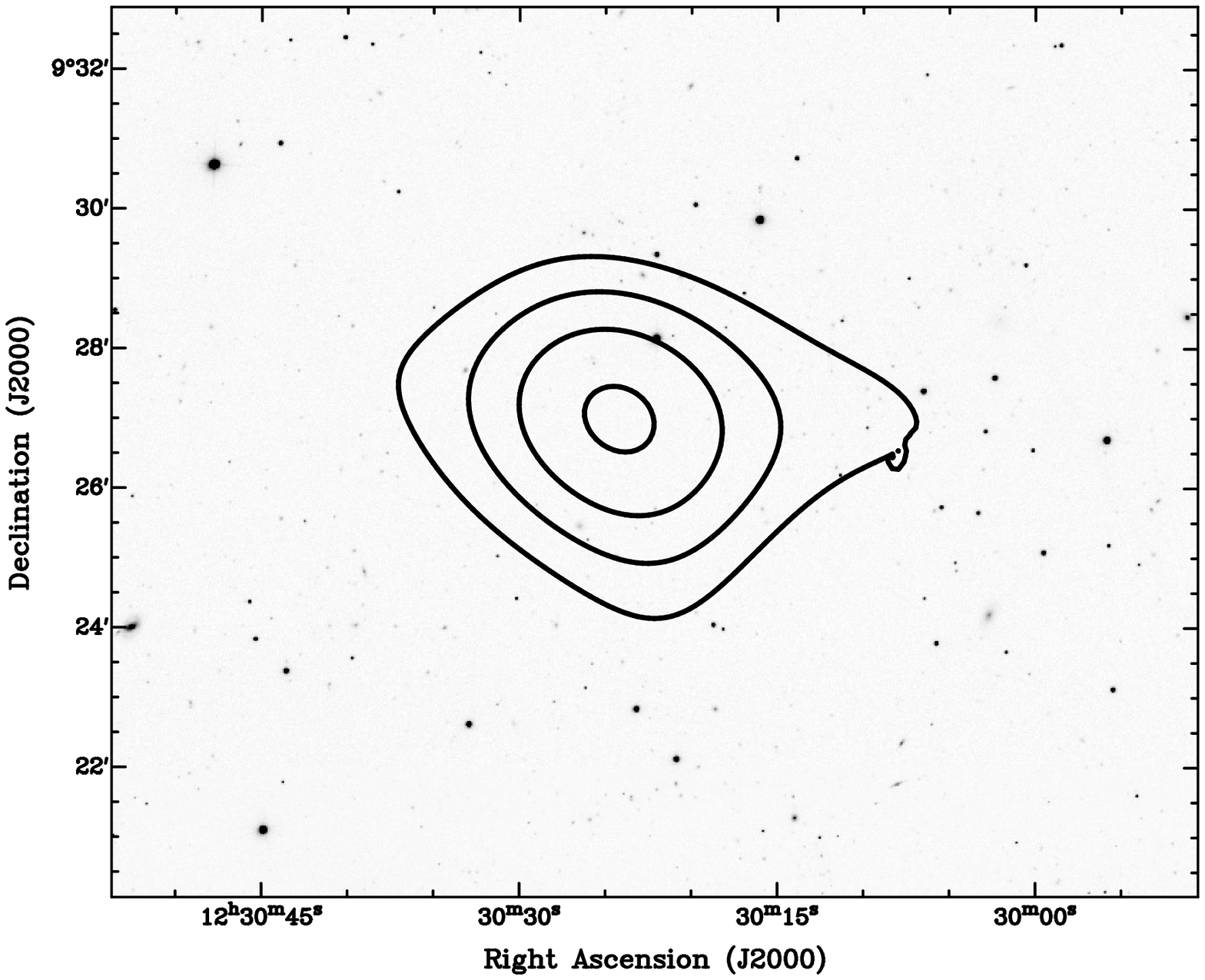}
	\includegraphics[width=0.6\textwidth]{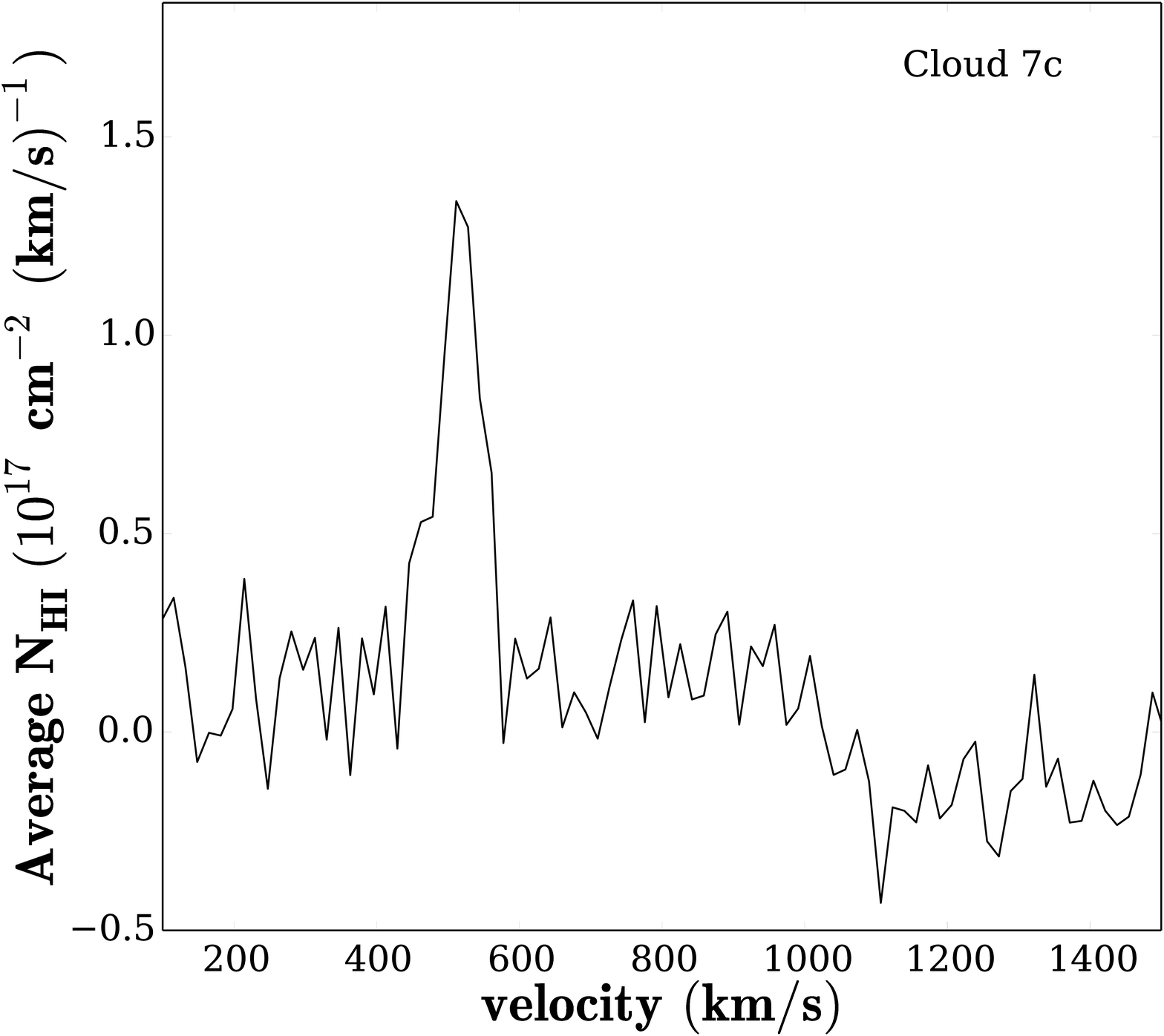}
	\caption{Column density map and \hi\, profile of Cloud 7c. Contours: 0.5, 1.0, 1.5, 2.0 $\times10^{19}\,\acm$. $v_{\rm sys} = 496$ \kms.}\label{app:cloud7c}
\end{figure}

\begin{figure}
	\centering
	\includegraphics[width=0.85\textwidth]{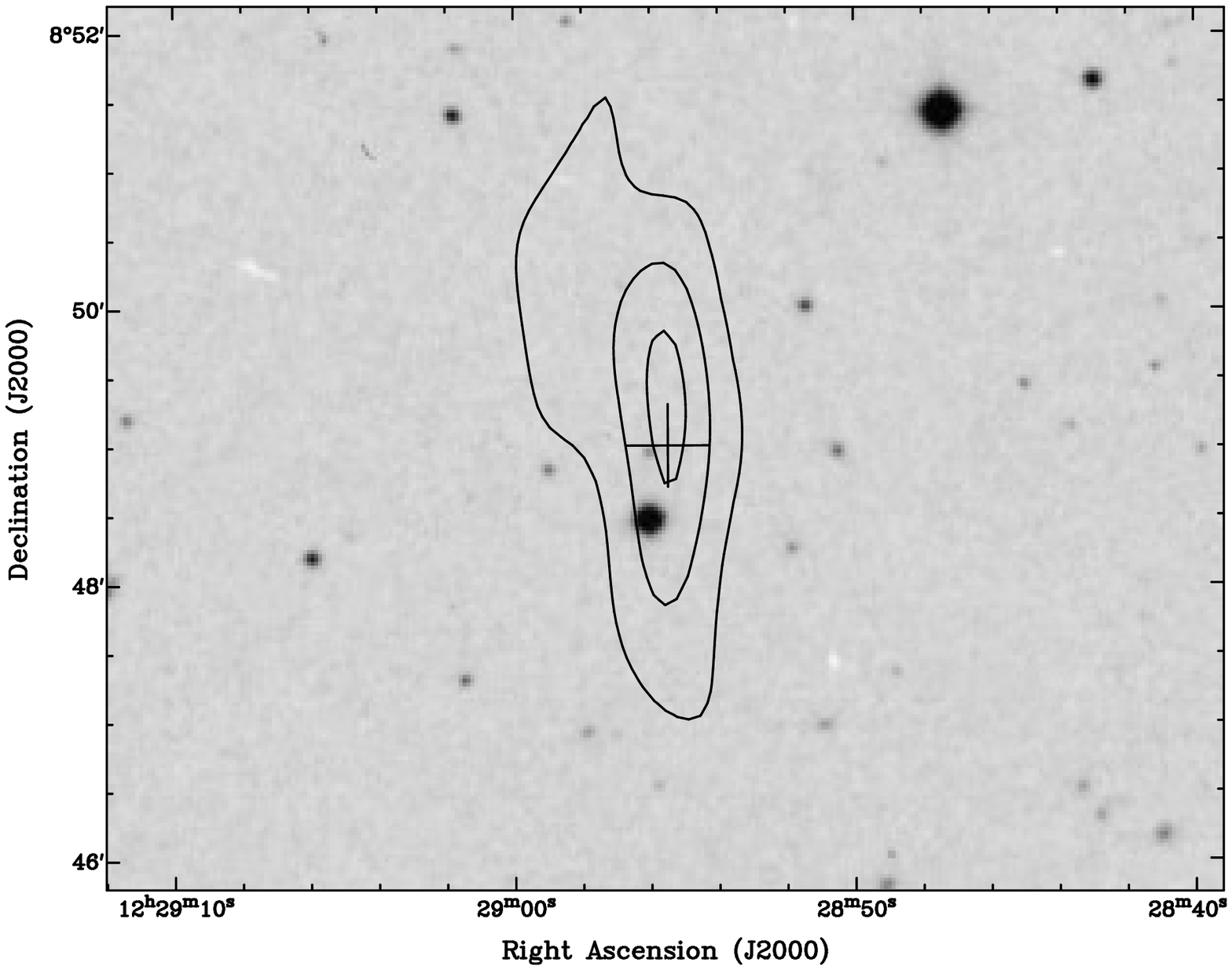}
	\includegraphics[width=0.6\textwidth]{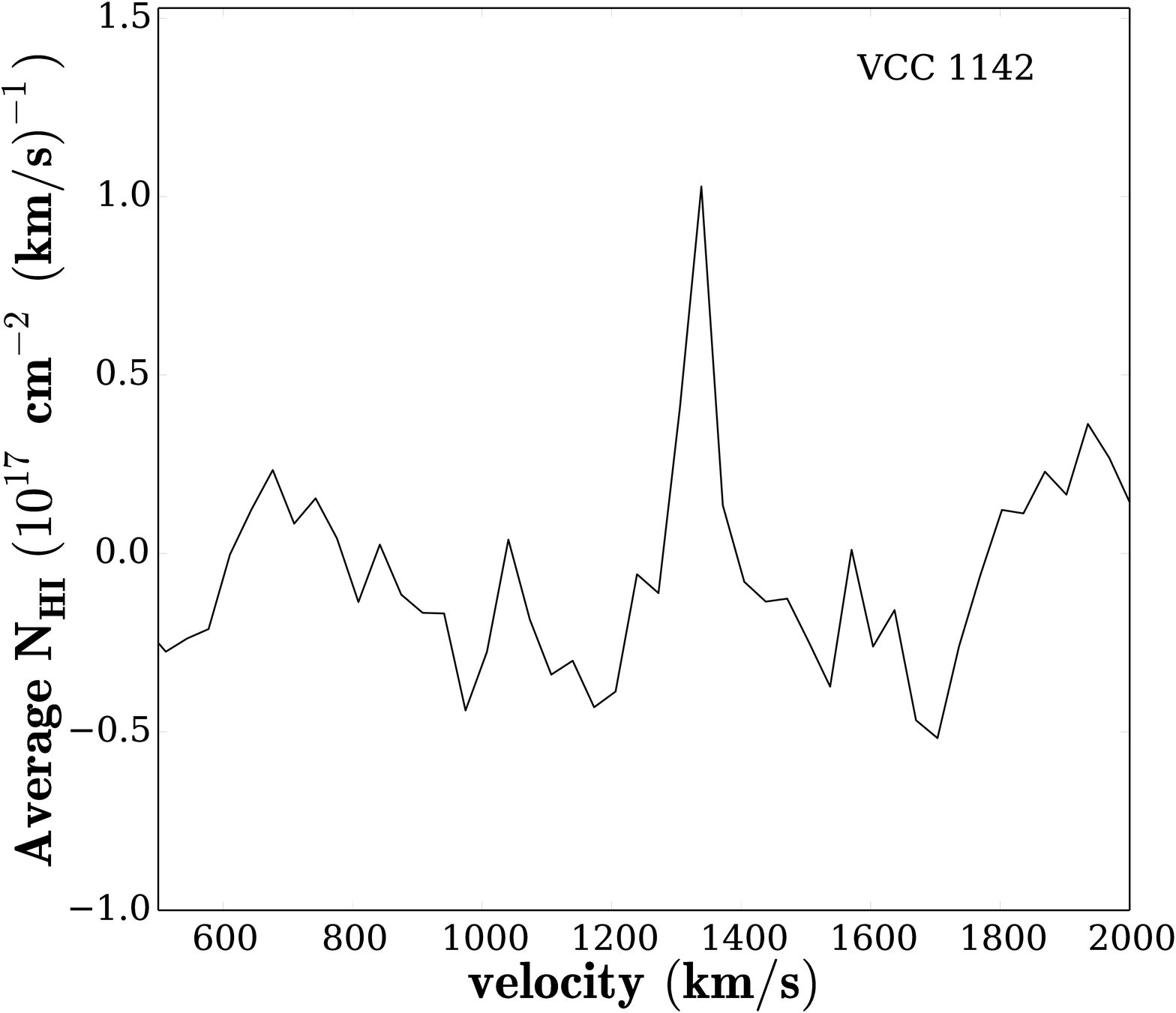}
	\caption{Column density map and \hi\, profile of VCC 1142. Contours: 0.5, 1.0, 1.5, 2.0 $\times10^{19}\,\acm$. $v_{\rm sys} = 1334$ \kms.}\label{app:v1142}
\end{figure}

\begin{figure}
	\centering
	\includegraphics[width=0.75\textwidth]{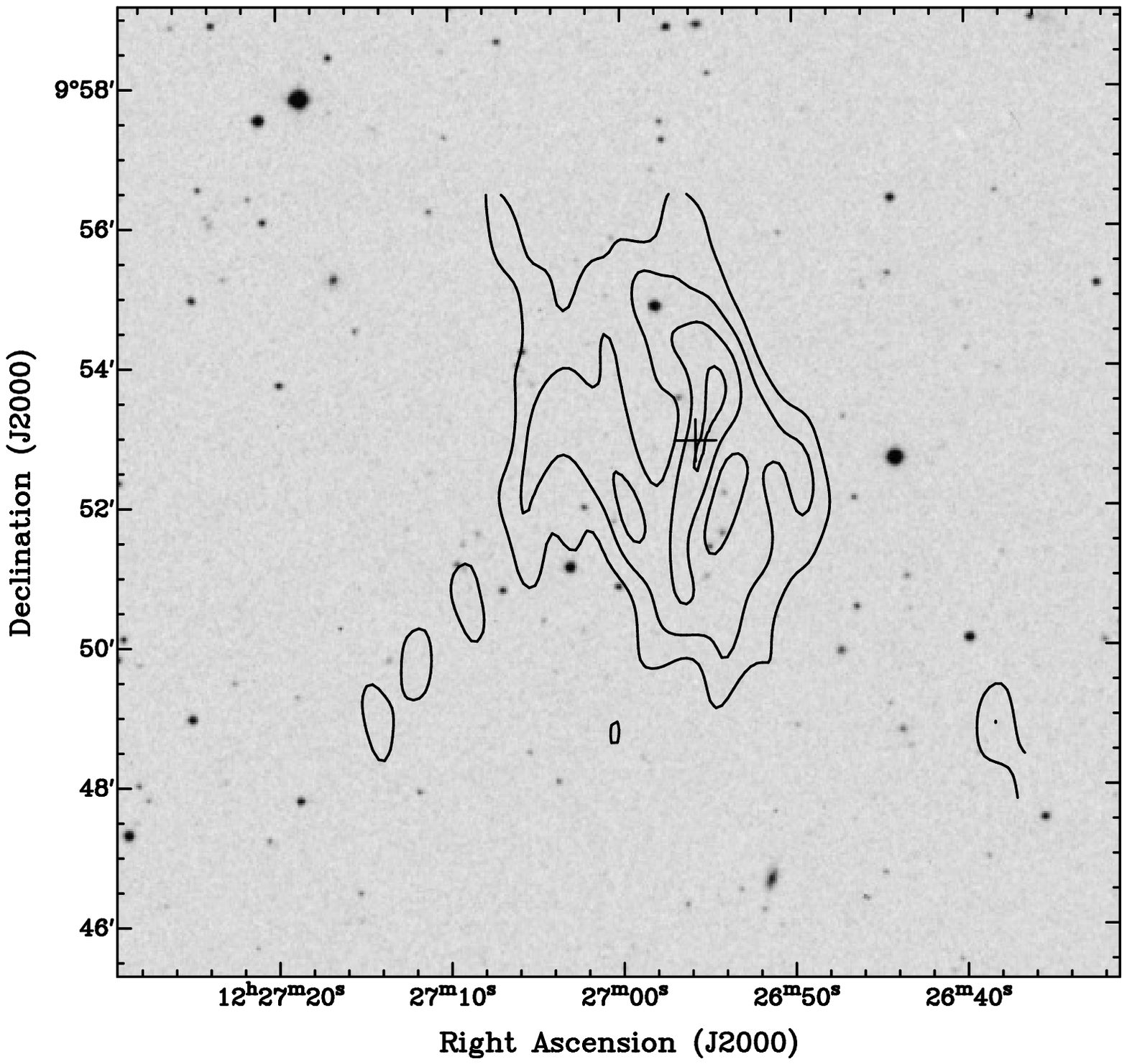}
	\includegraphics[width=0.6\textwidth]{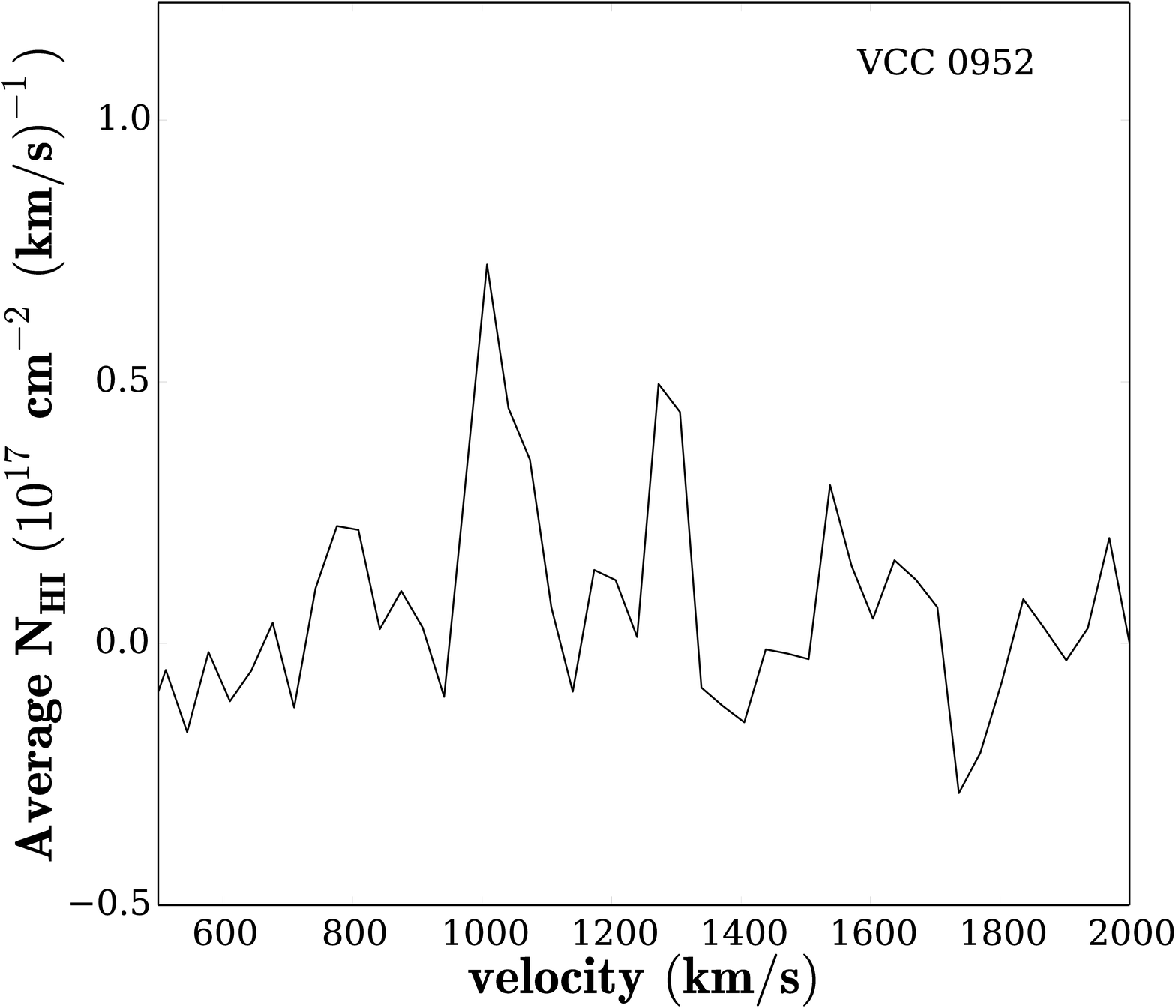}
	\caption{Column density map and \hi\, profile of VCC 0952. Contours: 0.5, 1.0, 1.5, 2.0 $\times10^{19}\,\acm$. $v_{\rm sys} = 1024$ \kms.}\label{app:v0952}
\end{figure}

\begin{figure}
	\centering
	\includegraphics[width=0.75\textwidth]{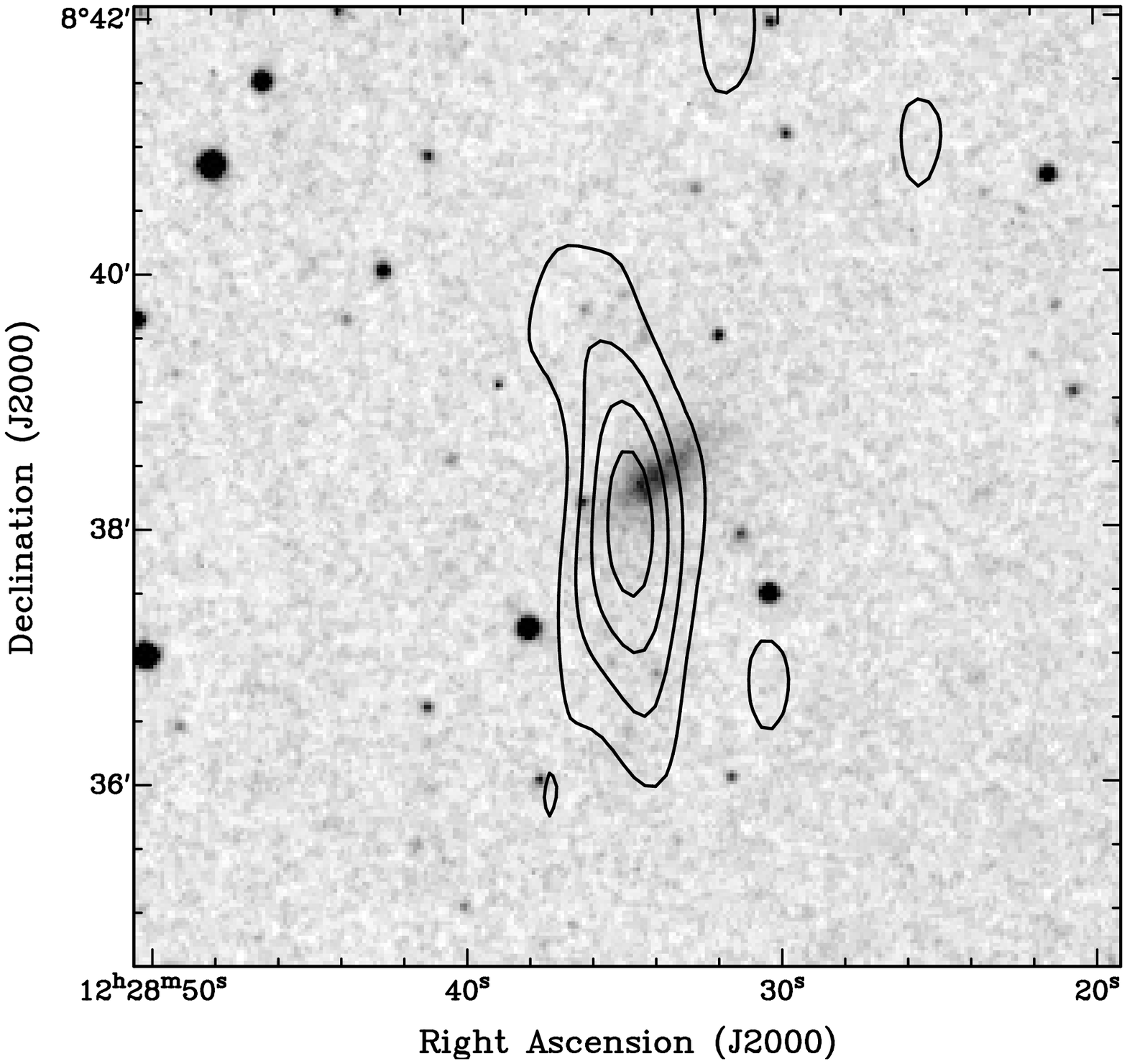}
	\includegraphics[width=0.6\textwidth]{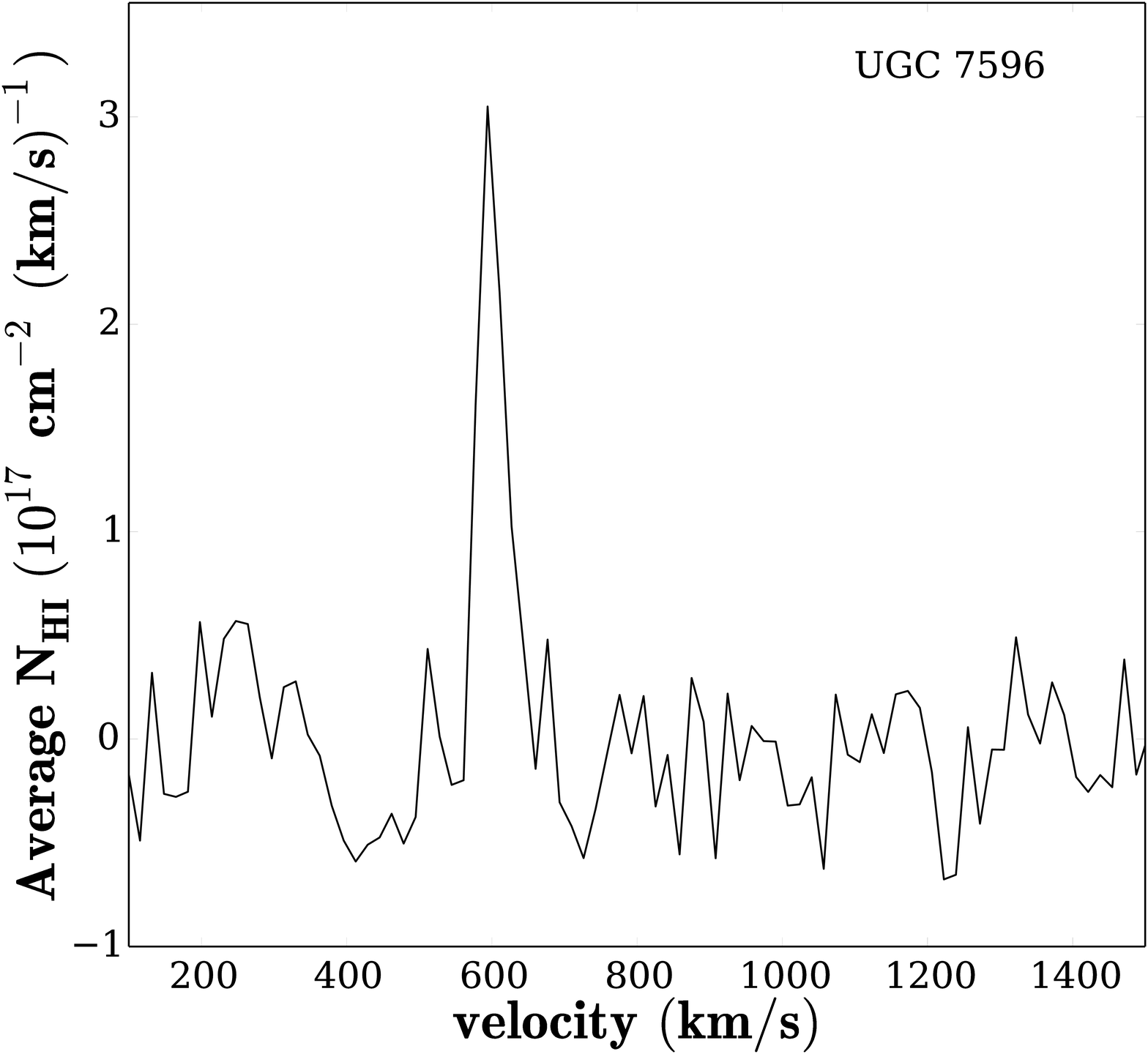}
	\caption{Column density map and \hi\, profile of UGC 7596. Contours: 0.5, 1.0, 1.5, 2.0 $\times10^{19}\,\acm$. $v_{\rm sys} = 595$ \kms.}\label{app:u7596}
\end{figure}

\begin{figure}
	\centering
	\includegraphics[width=0.75\textwidth]{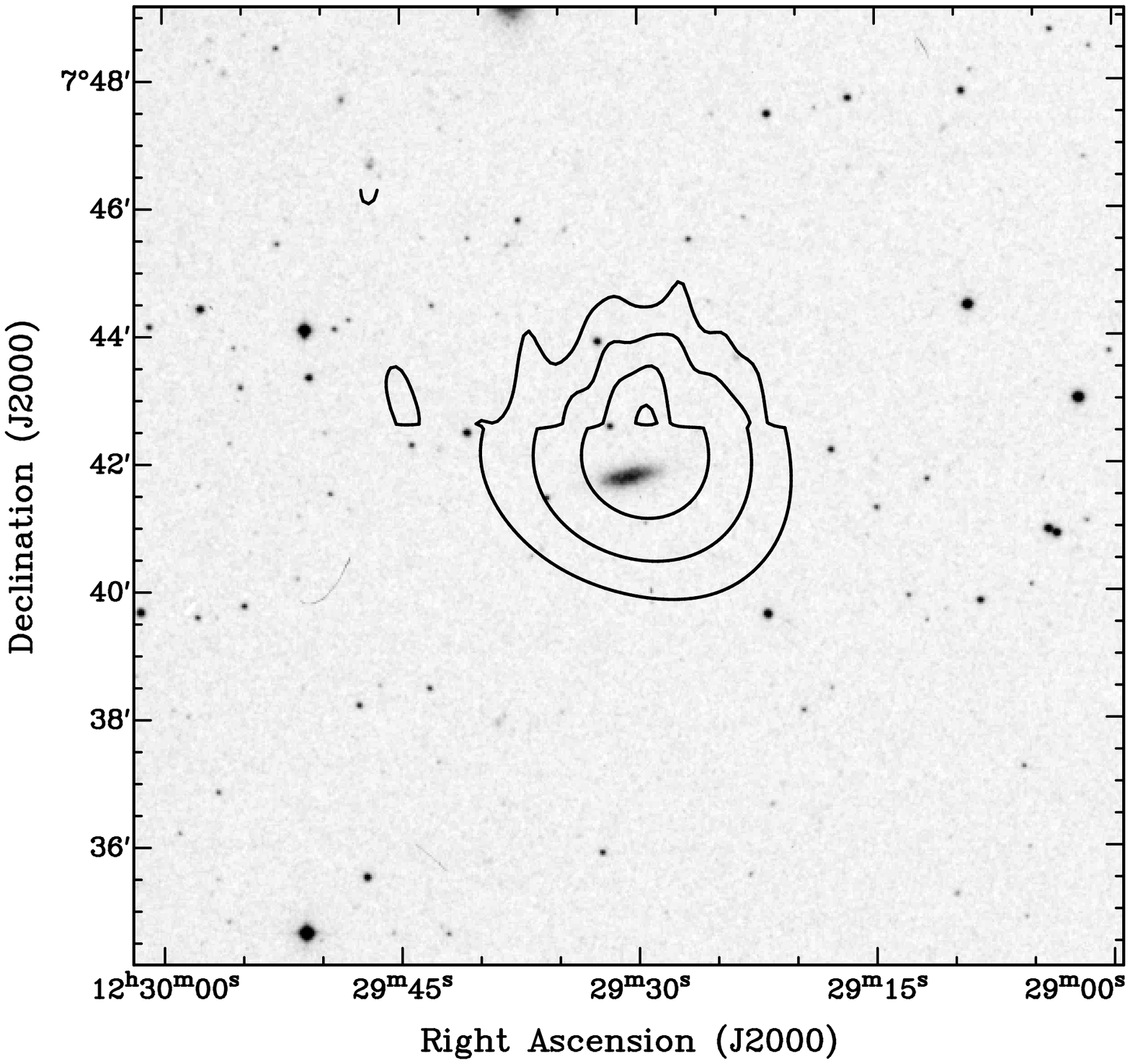}
	\includegraphics[width=0.6\textwidth]{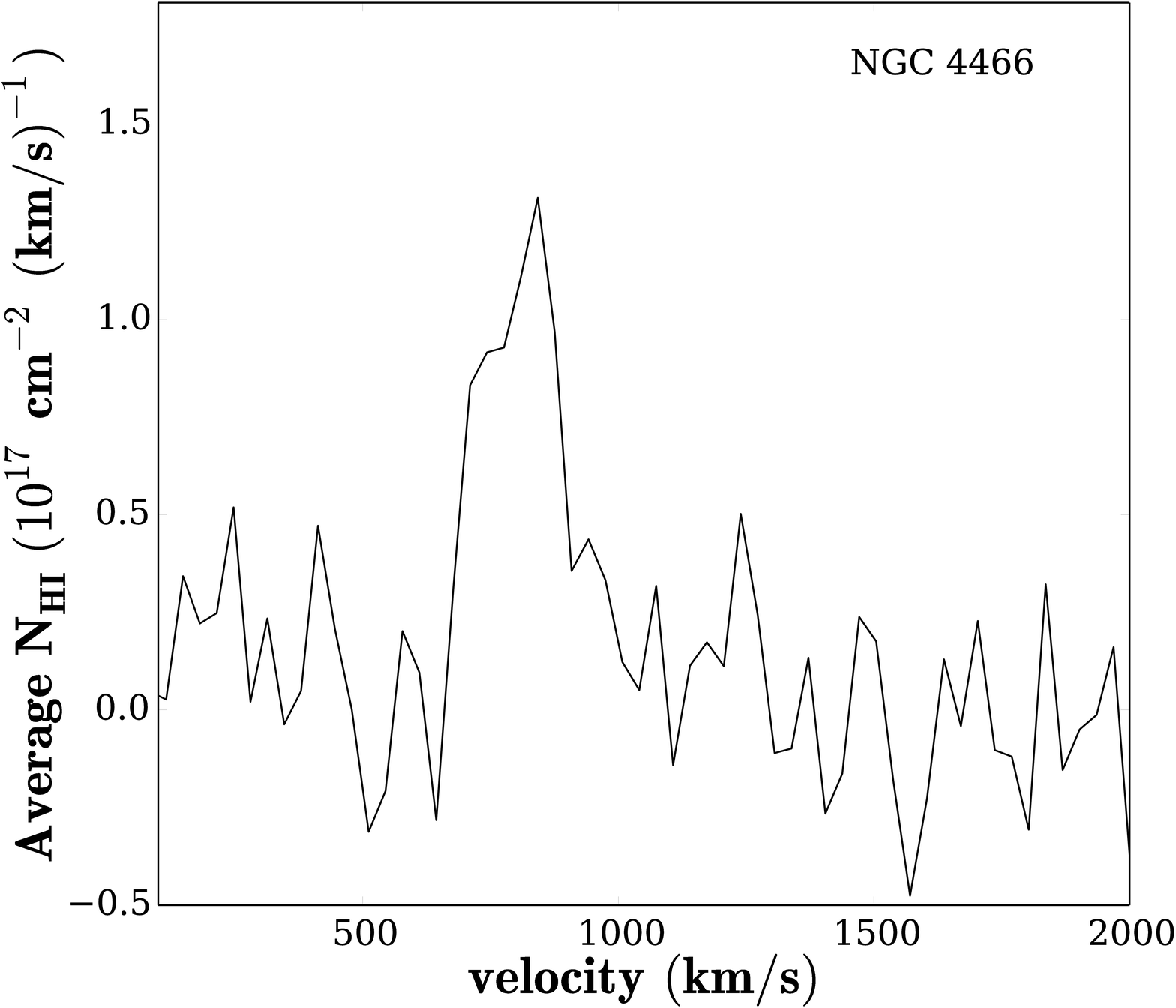}
	\caption{Column density map and \hi\, profile of NGC 4466. Contours: 1.0, 2.0, 3.0, 4.0 $\times10^{19}\,\acm$. $v_{\rm sys} = 797$ \kms.}\label{app:n4466}
\end{figure}

\begin{figure}
	\centering
	\includegraphics[width=0.75\textwidth]{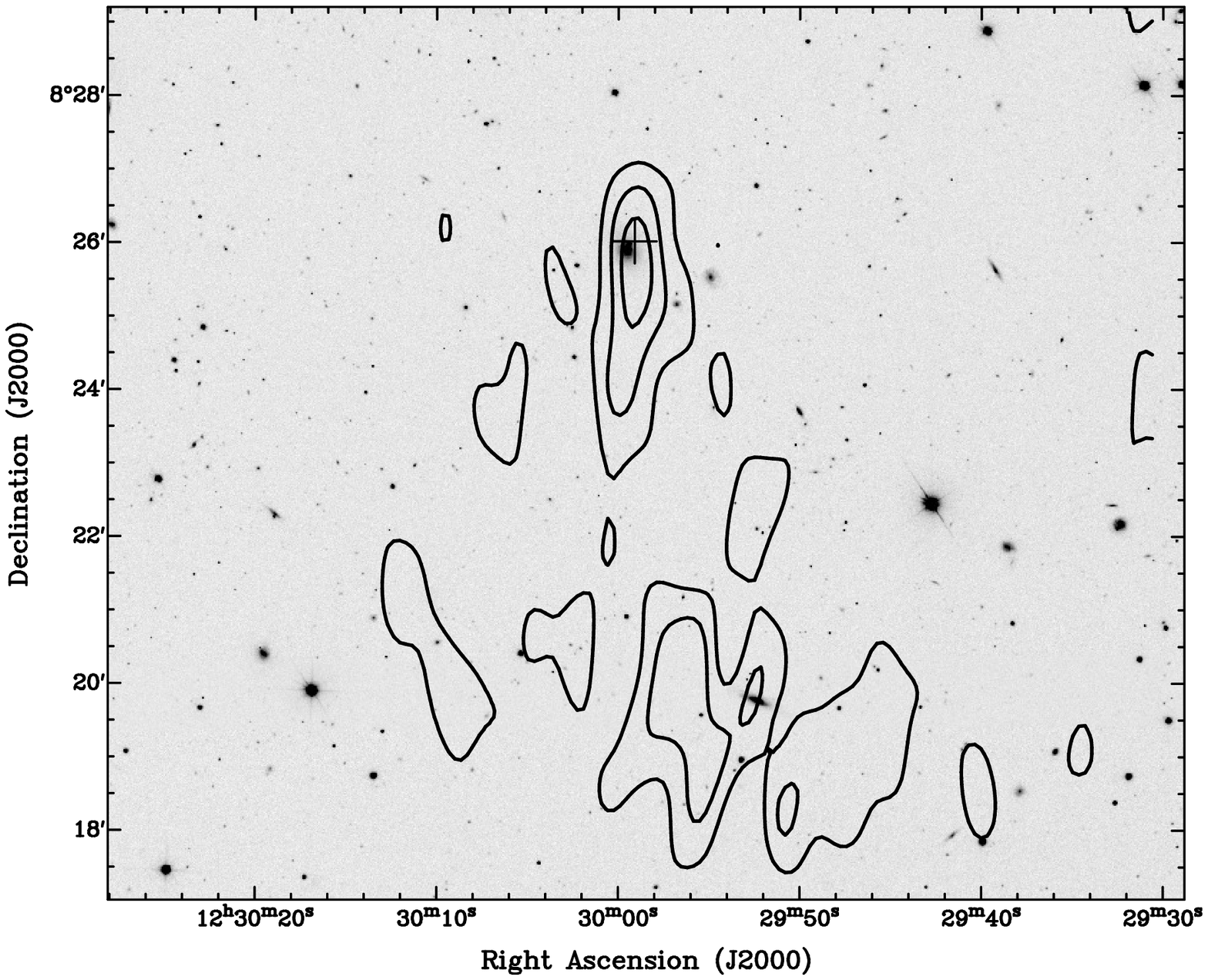}
	\includegraphics[width=0.6\textwidth]{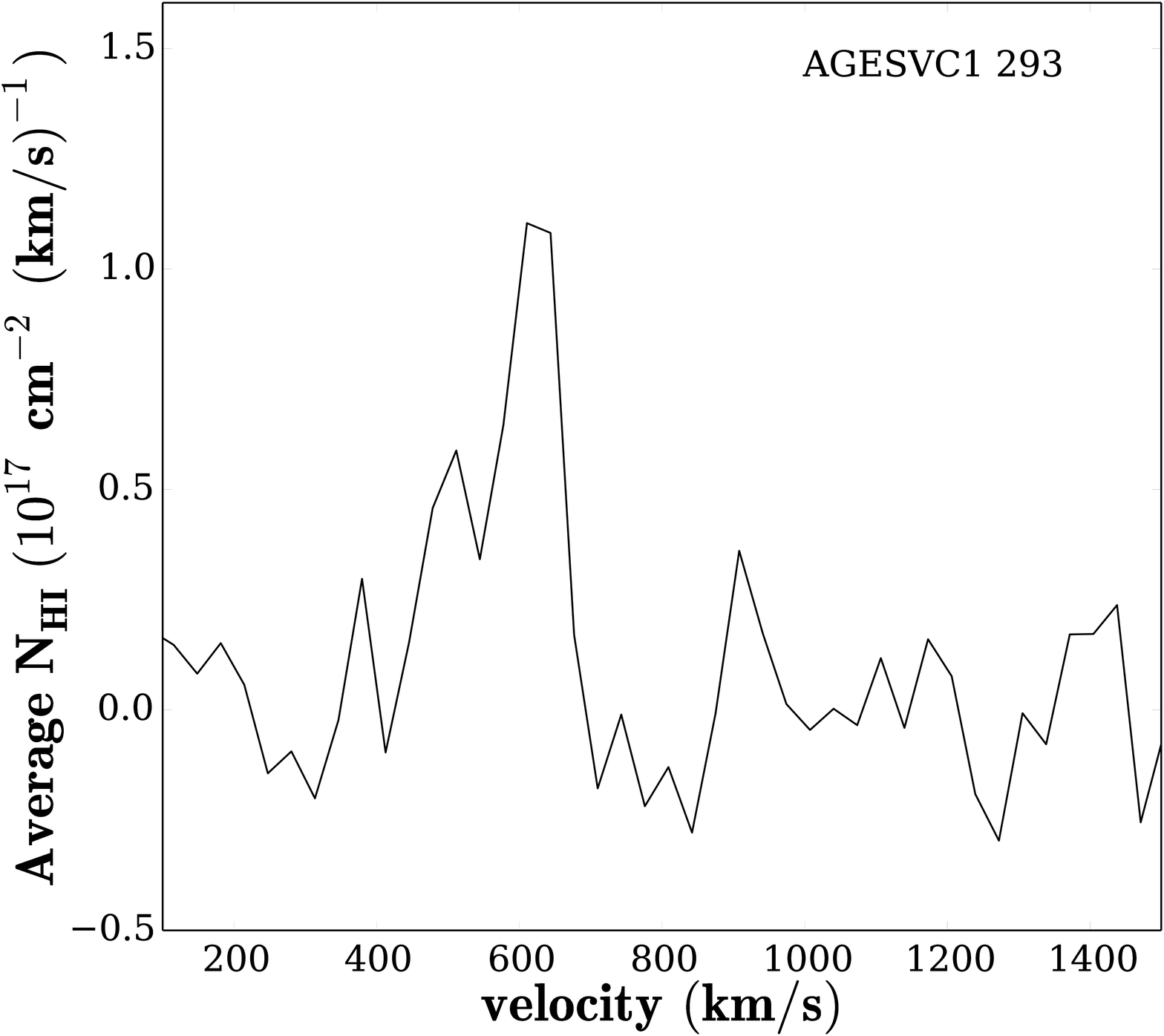}
	\caption{Column density map and \hi\, profile of AGESVC1 293. Contours: 1.0, 1.5, 2.0, 2.5 $\times10^{19}\,\acm$. $v_{\rm sys} = 595$ \kms. It is not clear whether the complex located south of the object (upper panel) is associated to the cloud, and further study may be needed to investigate this.}\label{app:ages293}
\end{figure}

\onecolumn
Below we present, from Figure \ref{app:n4424} to \ref{app:ages293}, the \hi\, column density maps of the detections along with their \hi\, profiles. For some of the galaxies, namely NGC 4424, NGC 4445, UGC 7557 and UGC 7596, the \hi\, centre is offset with respect to the optical centre. This offset is southwards, probable sign of environment effects.

Most of the objects lying in the south-east and north-west of the field (see Fig. \ref{fig:rosat-hi}) were not covered by the WSRT pointings, and therefore lack of resolution. NGC 4466 in particular, was partially covered and presents a northern half more resolved than the rest of the galaxy (Fig. \ref{app:n4466}).

\FloatBarrier

\clearpage
\thispagestyle{empty}

\clearpage
\thispagestyle{empty}

\clearpage
\thispagestyle{empty}

\end{document}